\documentclass[aps,prc,twocolumn,showpacs,preprintnumbers,
               nofootinbib,float,superscriptaddress,longbibliography]{revtex4-1}
\usepackage{graphicx, fancybox}
\usepackage{amsmath,amssymb}
\usepackage{color}
\usepackage[dvipsnames]{xcolor}
\usepackage[colorlinks=true, pdfstartview=FitV, linkcolor=Orange, citecolor=Cyan, urlcolor=blue]{hyperref}
\usepackage{soul}
\usepackage{array}
\usepackage{url}
\usepackage[utf8]{inputenc}
\newcommand{\ave}[1]{\left\langle #1 \right\rangle}

\newcommand{\snn}{\sqrt{s_\mathrm{NN}}}
\newcommand{\Rspin}{\mathcal{R}_{\Lambda}^{\hat{z}}}

\usepackage[normalem]{ulem}

\begin{document}

\title{Vortex Rings in Event-by-Event Relativistic Heavy-Ion Collisions}

\author{David Dobrigkeit Chinellato}
\affiliation{Instituto de Fisica Gleb Wataghin, Universidade Estadual de Campinas, Campinas, Brasil}

\author{Michael Annan Lisa}
\affiliation{Department of Physics, The Ohio State University, Columbus, Ohio, USA}

\author{Willian Matioli Serenone}
\affiliation{Instituto de Fisica Gleb Wataghin, Universidade Estadual de Campinas, Campinas, Brasil}

\author{Chun Shen}
\email{chunshen@wayne.edu}
\affiliation{Department of Physics and Astronomy, Wayne State University, Detroit, Michigan, 48201, USA}
\affiliation{RIKEN BNL Research Center, Brookhaven National Laboratory, Upton, NY 11973, USA}

\author{Jun Takahashi}
\affiliation{Instituto de Fisica Gleb Wataghin, Universidade Estadual de Campinas, Campinas, Brasil}

\author{Giorgio Torrieri}
\affiliation{Instituto de Fisica Gleb Wataghin, Universidade Estadual de Campinas, Campinas, Brasil}

\begin{abstract}
We present event-by-event simulations for central asymmetric light+heavy and Au+Au collisions to investigate the formation and evolution of vortex-ring structures in the longitudinal flow velocity profile. The production-plane polarization of $\Lambda$ hyperons, defined w.r.t. the $\Lambda$ momentum and the beam, can track the ``vortex-ring'' feature in the event, a characteristic vortical structure generated by longitudinal flow gradients. We make comprehensive model predictions for the rapidity-dependent vortex-ring observables for different collision system sizes at $\snn = 200$ and 72 GeV. Our predictions at the latter energy can be explored in the future LHCb fixed-target experiment at the Large Hadron Collider.
\end{abstract}

{\maketitle}

\section{Introduction}

One of the most common manifestations of hydrodynamics in everyday physics is the appearance of vortical structures originating from gradients~\cite{doi:10.1080/14786446708639824}. Most people are familiar with ``smoke rings'', in both air and water, generated by localized fast-moving currents embedded in a larger medium. Considering the seeming explanatory power of hydrodynamics in heavy ion collisions -- including small-on-large systems such as p+Au collisions --, it is interesting to look for vortex ring phenomena in such systems.

The appearance of collective phenomena in proton+nucleus collisions \cite{Dusling:2015gta, Nagle:2018nvi, Shen:2020mgh, Schenke:2021mxx, Noronha:2024dtq} makes these systems ideal laboratories for the search for such a phenomenon. 
The gradients produced by a ``bullet'' passing through a larger medium mimic the conditions associated with vortex ring formation. We have argued in Ref.~\cite{Lisa:2021zkj} that the production-plane polarization of $\Lambda$ hyperons can be used to measure the vortex-ring-like structure in the flow profile of produced deconfined matter (an analogous problem with jets was examined in \cite{jet1,jet2}). We initialized a three-dimensional viscous hydrodynamic simulation with a completely central collision (with impact parameter $b = 0$ fm) between a smooth proton and a smooth gold nucleus.

This work extends our simulations to more realistic conditions, including lumpy initial conditions and impact parameter fluctuations in event-by-event hydrodynamics. These simulations produce similar flow profiles; the vortex ring formation and observable are robust.
We present systematic studies on this vortex-ring observable for different light+heavy asymmetric collisions at $\sqrt{s_{\rm NN}}=200$ and 72 GeV. The system size scan allows us to study how the vortex ring observable approaches the limit of axially symmetric collision systems, such as Au+Au collisions. Our model predictions at $\snn = 72$\,GeV would motivate future measurements in the LHCb fixed-target experiments with the System for Measuring Overlap with Gas (SMOG) setup. The simulations at low collision energy also allow a quantitative comparison between our collective-dominated model and the earlier experimental results interpreted via QCD spin-orbit couplings~\cite{Heller:1977mv,Hauenstein:2016som,Bunce:1976yb,Agakishiev:2014kdy,Abt:2006da}.

Experimentally, the vortex ring structure can be quantified as~\cite{Lisa:2021zkj},
\begin{equation}
    \label{eq:RLambda3vectors}
   \mathcal{R}_{\Lambda}^{\hat{z}} \equiv
    2 \left\langle\frac{\vec{S}_{\Lambda} \cdot \left(\hat{z} \times \vec{p}_{\Lambda} \right)} {|\hat{z} \times \vec{p}_{\Lambda}|} \right\rangle_{\phi_\Lambda},
\end{equation}
where $\hat{z} \equiv (0, 0, 1)$ points in the direction of the light-ion beam, and the average is taken over the $\Lambda$ momentum azimuthal angle $\phi_\Lambda$ about the beam. The vectors $\vec{S}_{\Lambda}$ and $\vec{p}_{\Lambda}$ are the spin and momentum vectors for the $\Lambda$ hyperons in the lab frame, respectively.

The $\Lambda$'s spin polarization can be computed using the following the Cooper-Frye like formula from a vortical medium \cite{Becattini:2013fla, Becattini:2016gvu, Becattini:2020ngo}, 
\begin{equation}
  \label{eq:Freezeout}
   S^\mu(p)= - \frac{1}{8m} \epsilon^{\mu\rho\sigma\tau} \ave{\omega_{\sigma \rho} p_\tau},
\end{equation}
where $\omega_{\sigma \rho} \equiv -\frac{1}{2} [\partial_\sigma (u_\rho/T) - \partial_\rho (u_\sigma/T)]$ is the thermal vorticity tensor.
The $\ave{...}$ operator is defined as,
\begin{equation}
\label{eq:thermav}
\ave{X} \equiv 
    \frac{\int d\Sigma_\lambda p^\lambda n_F (1 -n_F) X}{\int d\Sigma_\lambda p^\lambda n_F},
\end{equation}
where $n_F$ is the Fermi-Dirac distribution, and $d\Sigma_\mu$ is the normal vector on the hydrodynamic freeze-out hypersurface.

Equation~\eqref{eq:Freezeout} builds the connection between the vortex ring observable $\mathcal{R}_{\Lambda}^{\hat{z}}$ with the production plane fluid vorticity density at freeze-out
\begin{equation}
\label{eq:Rfluid}
    \mathcal{R}_\mathrm{fluid}^{\hat{z}}
    \equiv 
    \frac{\epsilon^{\mu\nu\rho\sigma} \Omega_{\mu} n_{\nu} \hat{z}_{\rho} u_{\sigma}}
    {|\epsilon^{\mu\nu\rho\sigma}n_{\nu} \hat{z}_{\rho} u_{\sigma}|},
\end{equation}
where $u_\sigma$ is the fluid velocity and $\Omega^\mu \equiv \epsilon^{\mu\alpha\beta\gamma} \omega_{\alpha \beta} u_\gamma$ is the vorticity vector orthogonal to $u^\mu$. The unit vectors $\hat{z}^\rho = (0, 0, 0, 1)$ and $n^\mu$ points to the normal direction of the freeze-out surface, $n^\mu \equiv d\Sigma^\mu/|d\Sigma^\mu|$.

This paper will be laid out as follows. Section \ref{sec:model} will provide a concrete description of the (3+1)D dynamical model and computations of the $\Lambda$'s polarization vector. In Section \ref{sec:result}, we will discuss the ring observables in detail. We will conclude with some closing remarks in Sec.~\ref{sec:conc}.

In this paper, we use the conventions for the metric tensor $g^{\mu\nu} = \mathrm{diag}(1, -1, -1, -1)$ and the Levi-Civita symbol $\epsilon^{0123} = 1$.

\section{The Model Framework\label{sec:model}}

This work employs the geometric-based 3D initial conditions developed in Refs.~\cite{Shen:2020jwv, Ryu:2021lnx, Alzhrani:2022dpi} connecting with a hydrodynamics + hadronic transport model to carry out event-by-event simulations.
The transverse nuclear thickness functions for the two incoming nuclei are computed using
\begin{equation}
    T_{A(B)}(\vec{x}_\perp) = \sum_i \frac{1}{2\pi \omega^2} \exp\left(- \frac{|\vec{x}_\perp - \vec{x}_{\perp, i} |^2 }{2\omega^2} \right).
    \label{eq:TA}
\end{equation}
Here, the summation goes over all participant nucleons in the colliding nucleus. We assume a 2D Gaussian profile for each nucleon in the transverse plane with a width $\omega$ as a model parameter\footnote{The hot spot transverse size $w$ should not be confused with $\omega_{\mu \nu}$ and its modulus}.

We follow the 3D initial model \cite{Shen:2020jwv, Ryu:2021lnx, Alzhrani:2022dpi} to map the event-by-event nuclear thickness functions to the collision systems' initial energy-momentum tensor $T^{\mu\nu}$ at the starting time of hydrodynamic simulation $\tau = \tau_0 = 1$~fm/$c$. This initial-state model ensures the system's orbital angular momentum is conserved when mapping the energies and momenta of colliding particles to hydrodynamic fields. We assume the initial energy-momentum current takes the following form,
\begin{align}
    T^{\tau\tau} &= e(\vec{x}_\perp, \eta_s) \cosh(y_L(\vec{x}_\perp)), \label{eq:InitT00} \\
    T^{\tau\eta} &= \frac{1}{\tau_0} e(\vec{x}_\perp, \eta_s) \sinh(y_L(\vec{x}_\perp)). \label{eq:InitT03}
\end{align}
We assume there is no transverse flow at $\tau = \tau_0$, $T^{\tau x} = T^{\tau y} = 0$.
The system's longitudinal flow can be parameterized as
\begin{equation}
    y_L(\vec{x}_\perp) = f y_\mathrm{CM}(\vec{x}_\perp),
    \label{eq:yL}
\end{equation}
where the parameter $f \in [0, 1]$ controls how much of the initial net longitudinal momentum is attributed to the flow velocity. The $f = 0$ case recovers the well-known Bjorken flow profile, $y_L = 0$ in the Milne coordinates.
The center of mass rapidity $y_\mathrm{CM}$ is determined by the nuclear thickness functions in every transverse position,
\begin{equation}
    y_\mathrm{CM}(\vec{x}_\perp) = \mathrm{arctanh}\left[\frac{T_A - T_B}{T_A + T_B} \tanh(y_\mathrm{beam}) \right].
    \label{eq:yCM}
\end{equation}

The local energy density in Eqs.~\eqref{eq:InitT00} and \eqref{eq:InitT03} is parametrized as 
\begin{align}
    & e (\vec{x}_\perp, \eta_s, y_\mathrm{CM},f) = \nonumber \\
    & \quad \mathcal{N}_e(\vec{x}_\perp) \exp\bigg[- \frac{(\vert \eta_s - (y_\mathrm{CM}(1 - f)) \vert  - \eta_0)^2}{2\sigma_\eta^2} \nonumber \\
    & \quad \qquad \qquad \qquad \times \theta(\vert \eta_s - (y_\mathrm{CM} (1- f)) \vert - \eta_0)\bigg],
    \label{eq:eprof}
\end{align}
where $\eta_s$ is the spacetime rapidity and $f$ parametrizes the radial gradient of the longitudinal flow (see \cite{Shen:2020jwv} for a discussion). As shown, $f=0$ implies everything depends on $\eta-y_\mathrm{CM}$. The normalization factor $\mathcal{N}_e(\vec{x}_\perp)$ is fixed by the total incoming collision energy, and it scales with $\sqrt{T_A(\vec{x}_\perp) T_B(\vec{x}_\perp)}$ at high energies~\cite{Shen:2020jwv}. The values of the model parameters are listed in Table~\ref{tab:model_params}. The initial baryon density distribution is set up as in Ref.~\cite{Shen:2020jwv}.

The initial energy-momentum tensor is propagated hydrodynamically with lattice QCD EoS~\cite{Monnai:2019hkn} using the numerical code \texttt{MUSIC} \cite{Schenke:2010nt, Schenke:2010rr, Paquet:2015lta, Denicol:2018wdp}, based on the Denicol-Niemi-Molnar-Rischke (DNMR) hydrodynamic model~\cite{Denicol:2012cn, Denicol:2018wdp}. This work only considers shear viscous effects and uses a constant QGP specific shear viscosity $\eta T/(e + P) = 0.08$. 
Particlization is performed on a constant energy density hyper-surface with $e_{\rm sw} = 0.5$~GeV/fm$^3$ identified by the \texttt{Cornelius} algorithm~\cite{Huovinen:2012is}.  Individual hyper-surface cells contain all the ingredients needed to calculate the ring polarization observable according to Eq.~\eqref{eq:Freezeout}. We will explore the $\Rspin$ dependence on model parameters in Sec.~\ref{sec:modelDep}. The numerical simulations are carried out using the \texttt{iEBE-MUSIC} framework. 

Since the ring observable in Eq.~\eqref{eq:RLambda3vectors} is driven by longitudinal velocity and density gradients, both the hotspot transverse size $w$ and $f$ are expected to influence it. As already discussed in \cite{Ryu:2021lnx, Alzhrani:2022dpi}, the vortical structure is sensitively related to $f$ because deviation from the limit of absolute transparency in the Bjorken picture also quenches longitudinal vorticity. However, longitudinal vorticity is also quantitatively driven by gradients sensitive to the transverse in-homogeneity scale, which, in the Glauber model, is determined by the nuclear size \cite{Alzhrani:2022dpi}.

\begin{table}[h!]
    \centering
    \caption{Model parameters used in the simulations.}
    \begin{tabular}{c|c|c}
    \hline \hline
       Model Parameter  &  $\snn = 200$~GeV  & $\snn = 72$~GeV \\ \hline
       $\eta_0$  & 5.0  &  4.5 \\ \hline
       $\sigma_\eta$ & 0.55 & 0.5 \\ \hline
    \hline
    \end{tabular}
    \label{tab:model_params}
\end{table}

\begin{figure}[h!]
  \centering
  \includegraphics[width=1.0\linewidth]{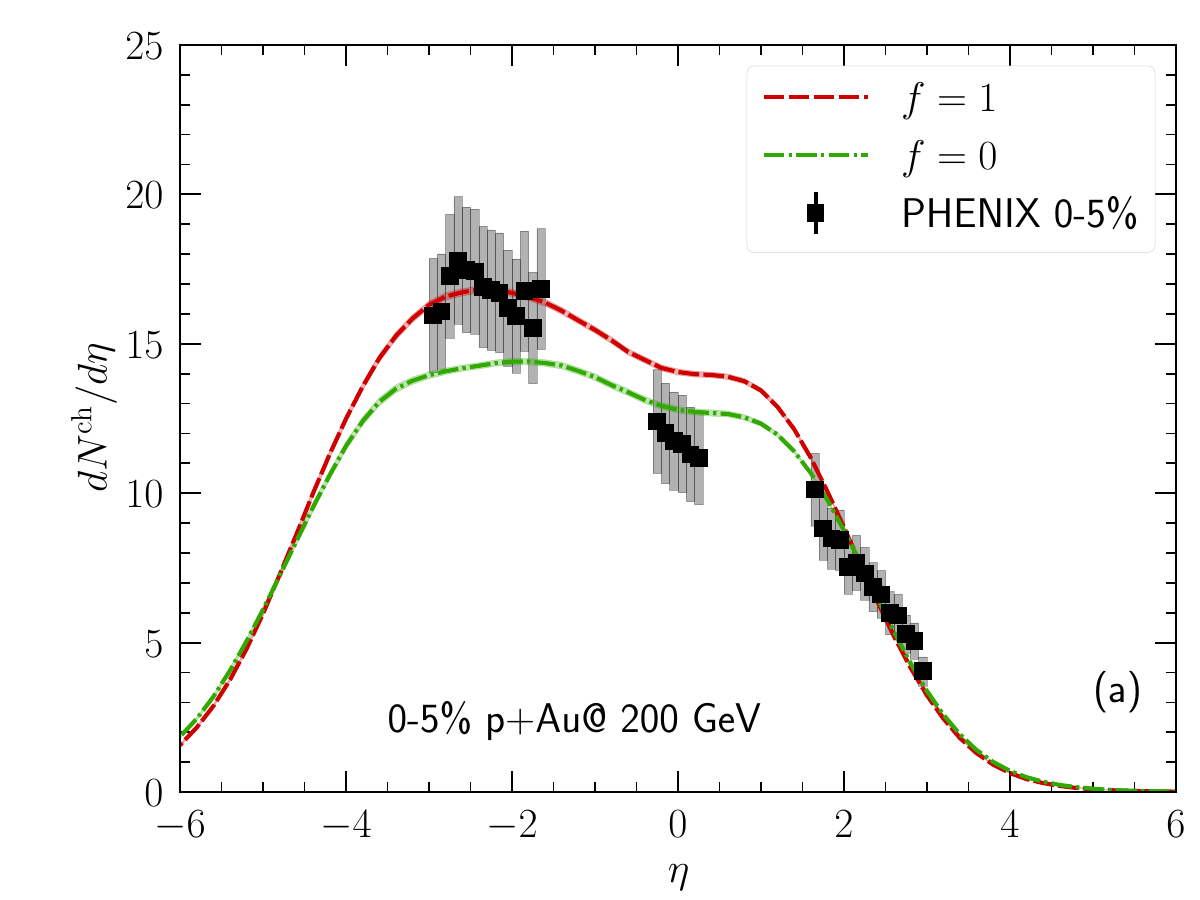}
  \includegraphics[width=1.0\linewidth]{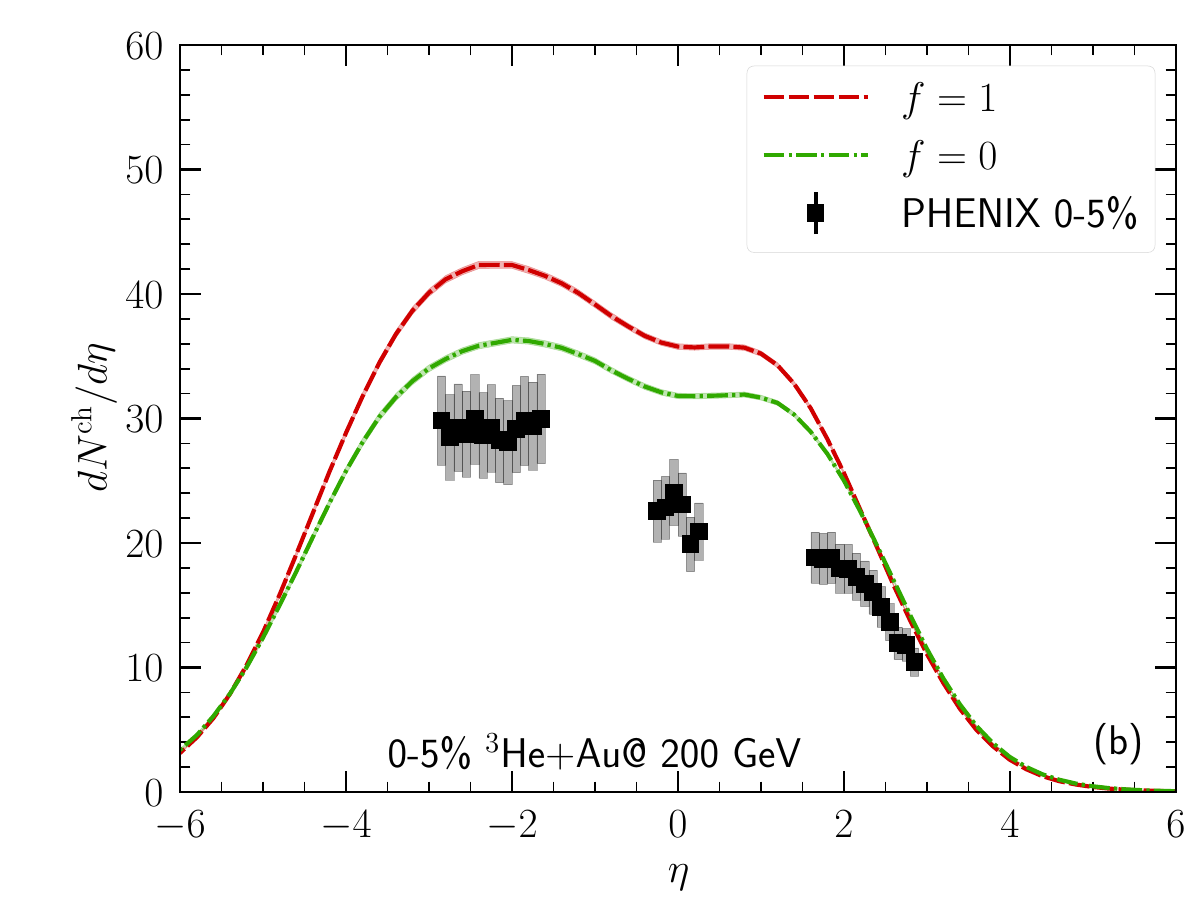}
  \includegraphics[width=1.0\linewidth]{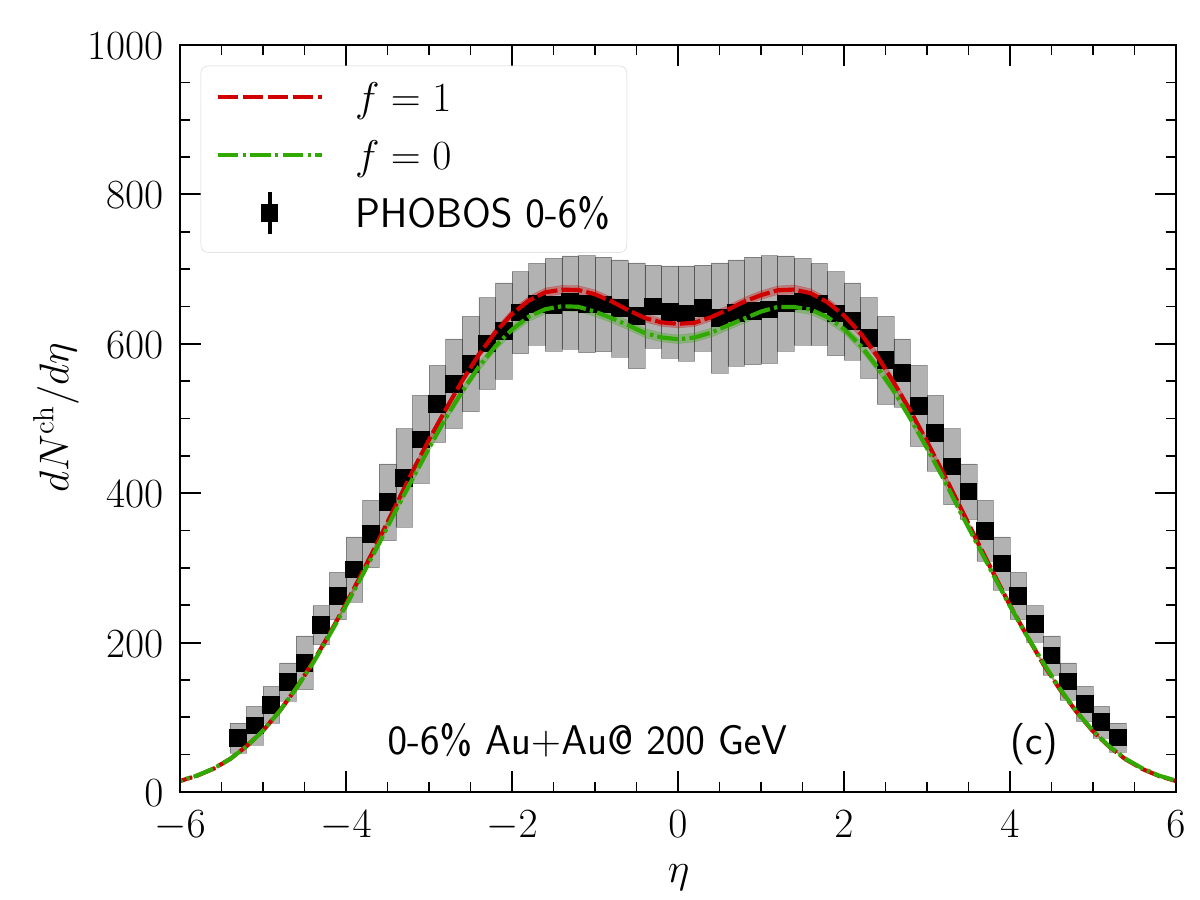}
  \caption{Charged hadron pseudorapidity distributions for 0-5\% p+Au, $^{3}$He+Au, Au+Au collisions at 200 GeV with different values of model parameter $f$ compared with PHENIX and PHOBOS measurements~\cite{PHENIX:2018hho, PHOBOS:2005zhy}.}
  \label{fig:dNdeta}
\end{figure}

\section{Results}\label{sec:result}

\subsection{Vortex rings at the top RHIC energy}

First, we will calibrate our model to some experimental measurements that characterize the global properties of the collision systems. The charged hadron $dN/d\eta$ is shown in Fig.~\ref{fig:dNdeta} for different collision systems at $\snn = 200$~GeV.
Our calculations give a qualitative agreement to the PHENIX and PHOBOS measurements~\cite{PHENIX:2018hho, PHOBOS:2005zhy}. However, quantitatively, the asymmetry is smaller in this initial-state model than in the PHENIX measurements for small systems. We note that the global longitudinal multiplicity is dependent weakly on model parameter $f$, which controls the magnitude of the initial longitudinal flow.

\begin{figure}[b!]
  \centering
  \includegraphics[width=1.0\linewidth]{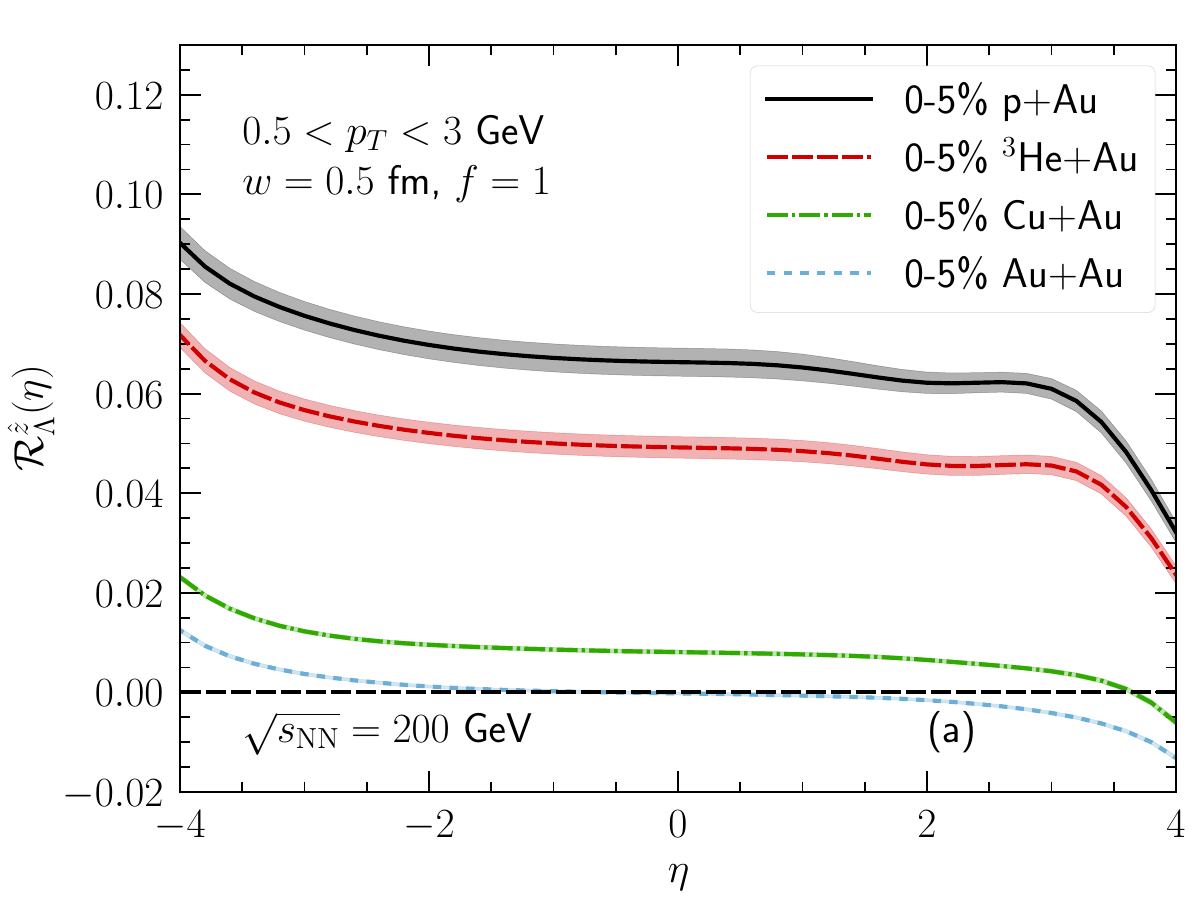}
  \includegraphics[width=1.0\linewidth]{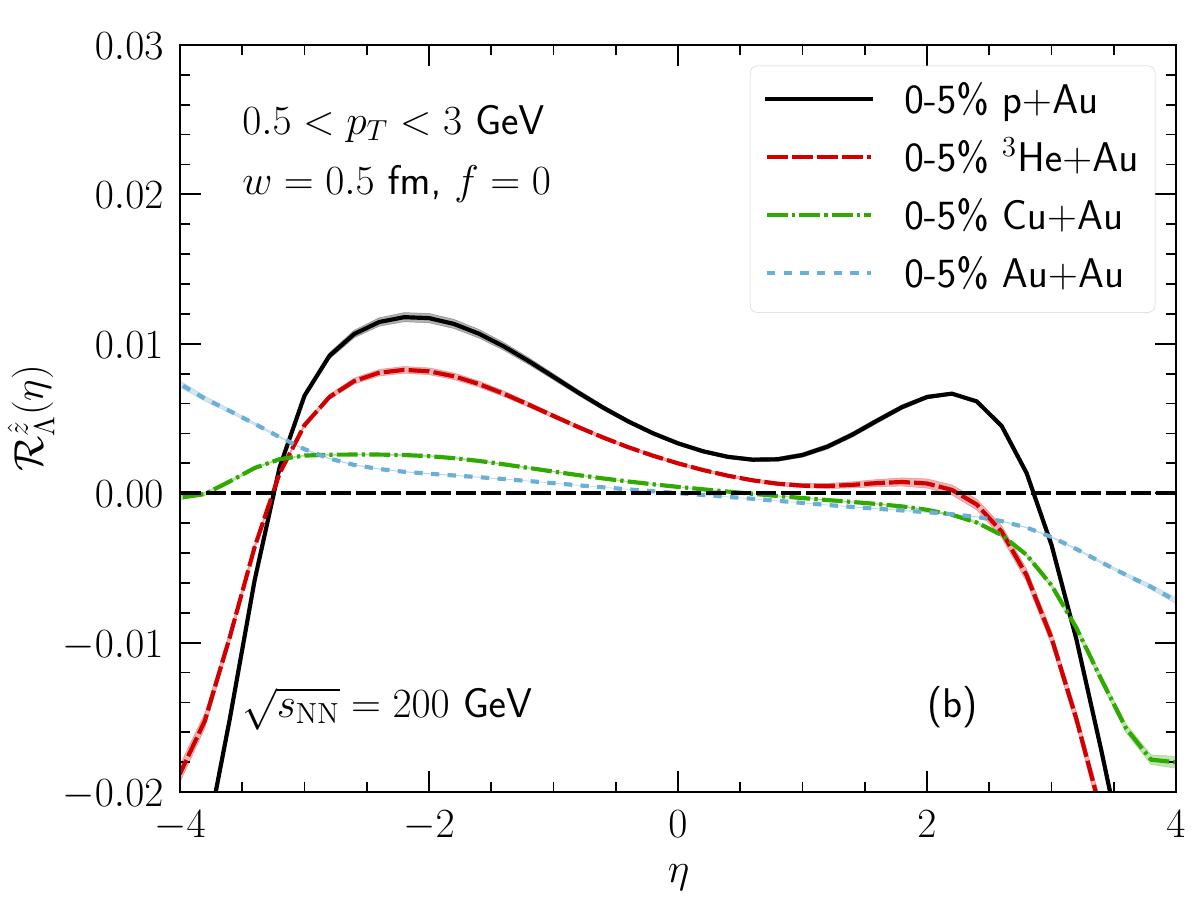}
  \caption{The collision system size dependence of the hyperon's $\Rspin(\eta)$ in central light+heavy ion collisions at $\snn = 200$\,GeV. Two values of $f = 0, 1$ are shown in panels (a) and (b).}
  \label{fig:RspinSys}
\end{figure}

Figure~\ref{fig:RspinSys} shows the $\Rspin$ observable (defined in Eq.~\eqref{eq:RLambda3vectors}) as a function of $\Lambda$'s pseudo-rapidity for several collision systems at $\snn = 200$~GeV. Comparing panels (a) and (b), we find that the observable $\Rspin$ shows a strong sensitivity to the amount of initial-state longitudinal flow used in the model. The sensitivity increases as the collision systems become more and more asymmetric. The values of $\Rspin$ are about a factor of 10 different in the two sets of simulations for central p+Au collisions. Therefore, a measurement of $\Rspin$ can serve as a direct probe for the system's initial-state longitudinal flow in asymmetric collisions. A collision system scan of this observable, as shown in Fig.~\ref{fig:RspinSys}, would be a valuable tool to reveal the early-stage stopping dynamics in relativistic heavy-ion collisions.

\begin{figure}[t!]
  \centering
  \includegraphics[width=1.0\linewidth]{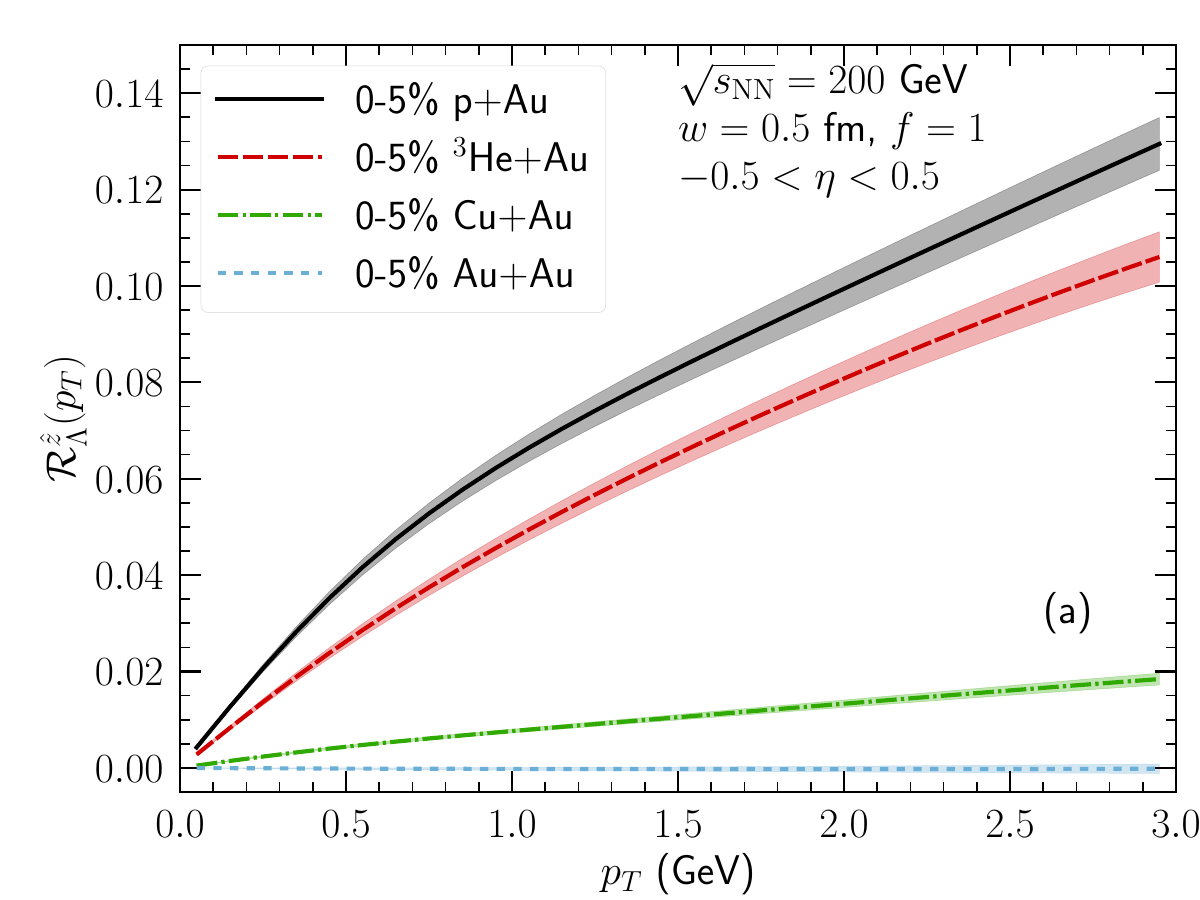}
  \includegraphics[width=1.0\linewidth]{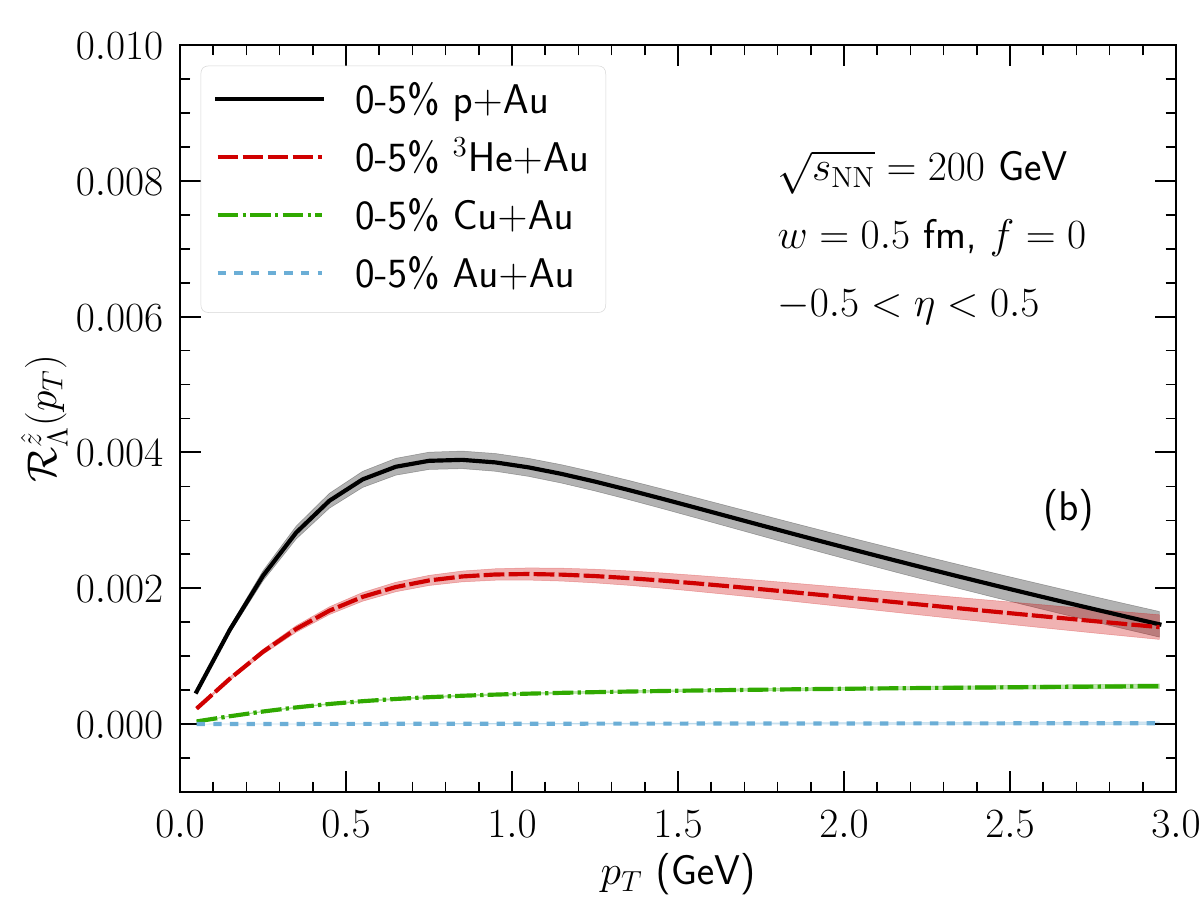}
  \caption{The $p_T$-differential hyperon's $\Rspin(p_T)$ in central heavy-ion collisions at $\snn = 200$\,GeV. Two values of $f$ are shown.}
  \label{fig:RspinpT_sys}
\end{figure}
\begin{figure}[t!]
  \centering
  \includegraphics[width=1.0\linewidth]{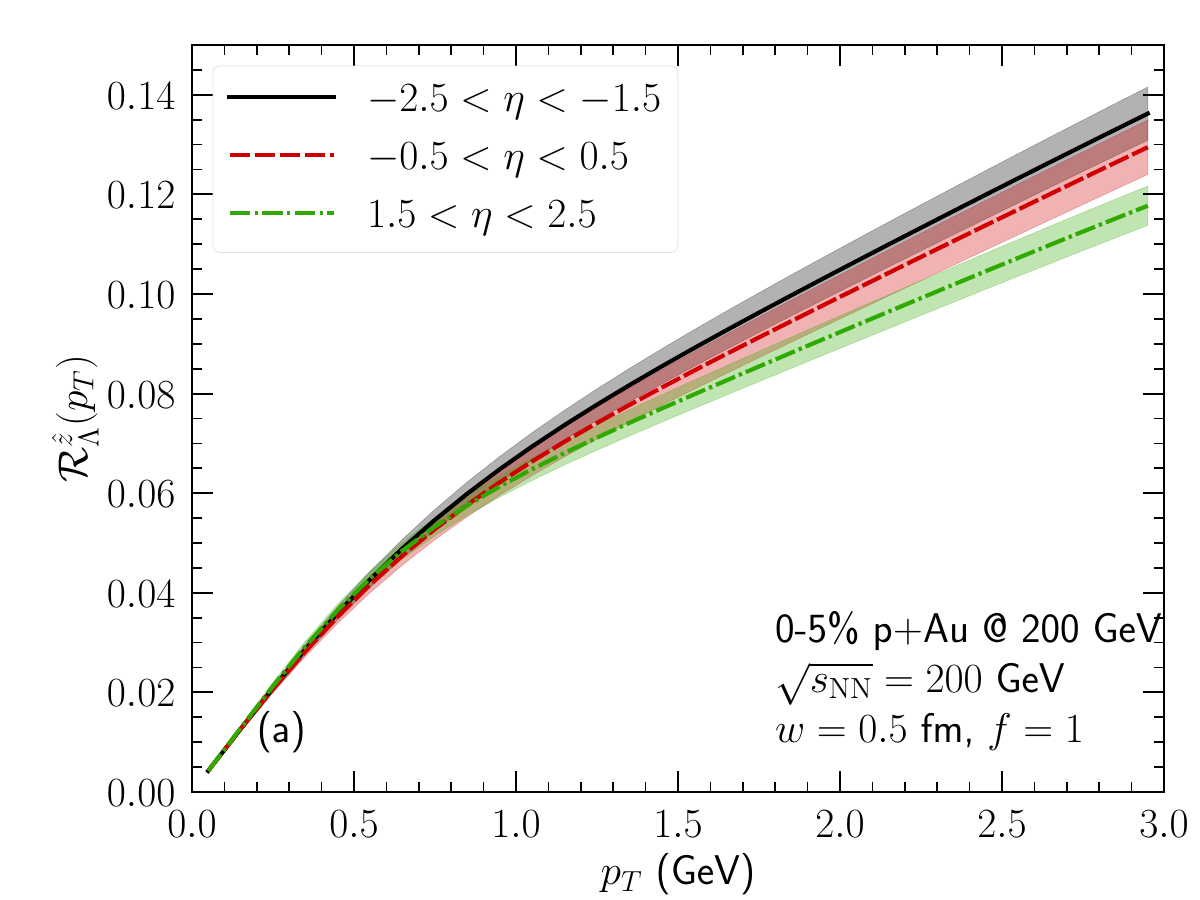}
  \includegraphics[width=1.0\linewidth]{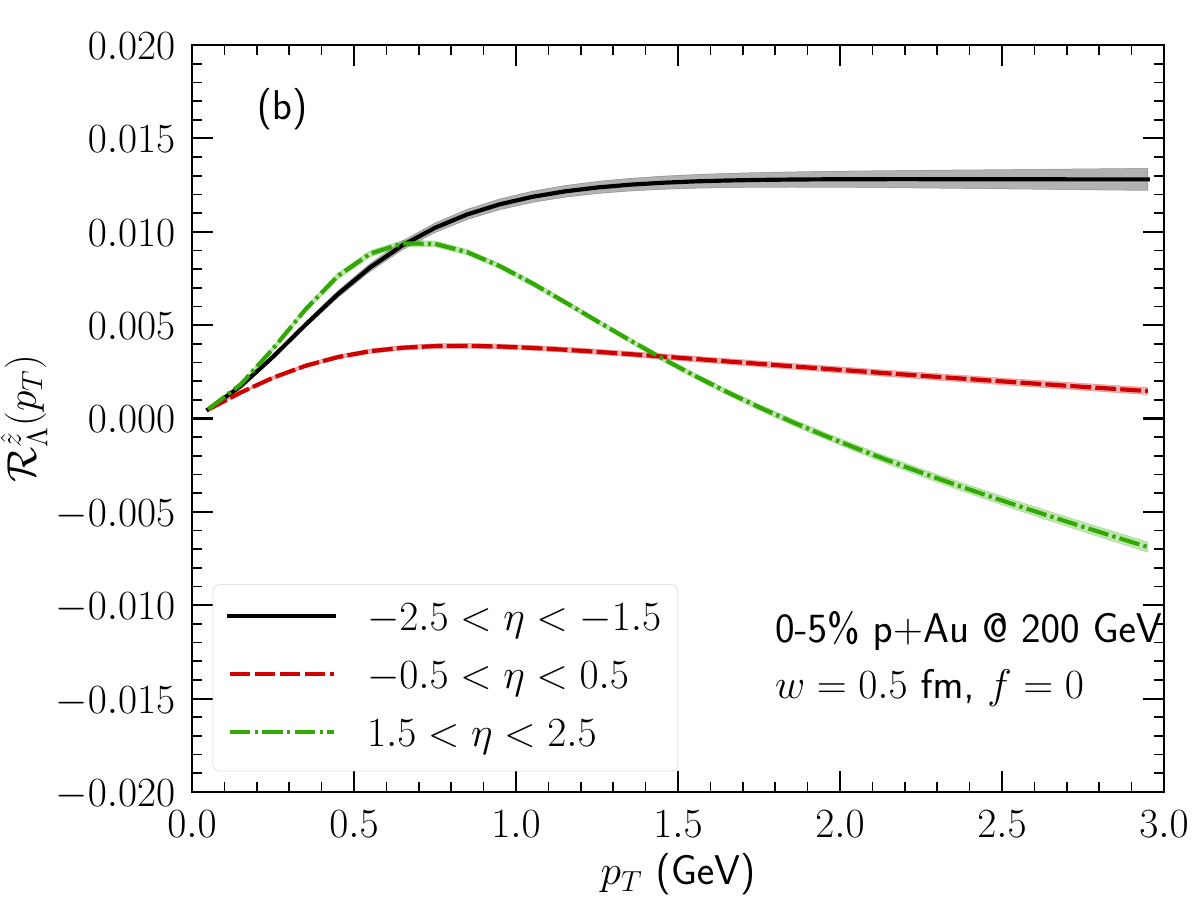}
  \caption{The $p_T$-differential hyperon's $\Rspin(p_T)$ in central p+Au collisions at $\snn = 200$\,GeV. Two values of $f$ are shown.}
  \label{fig:RspinpT_pAu_rap}
\end{figure}

With a strong initial-state longitudinal flow ($f = 1$), the $\Rspin(\eta)$ imprints the early-stage flow vorticity pattern from the initial collision configurations. The approximately constant $\Rspin$ within $\vert \eta \vert < 3$ shows the light ion drills through the heavy nucleus. For the scenario with $f = 0$, the collision systems develop fluid vorticity from zero by the local pressure gradients. Such hydrodynamic response to the geometry develops fluid vorticity slowly, resulting in a smaller effect on the $\Lambda$'s polarization.

The rapidity-odd $\Rspin(\eta)$ in Au+Au collisions is a signature of the fireball's transverse expansion~\cite{Lisa:2021zkj}. As the collision systems become more asymmetric, the values of $\Rspin$ stay positive within $\vert \eta \vert < 3$ for both sets of simulations ($f = 0, 1$). These results suggest that the longitudinal expansion away from the Bjorken flow is stronger than the fireball's transverse expansion in small systems, such as p+Au and $^3$He+Au collisions.

Now, we would like to understand the transverse momentum $p_T$ dependence of the $\Rspin$ observable.
Based on Eqs.~\eqref{eq:RLambda3vectors} and \eqref{eq:Freezeout}, we can show that 
\begin{align}
     \Rspin(p_T) &\propto \ave{\frac{1}{m}(E \omega^{z \sigma}  - p^z \omega^{t \sigma})\frac{p_\sigma}{p_T} + \frac{m}{p_T} \omega^{tz}} \notag \\
     &\propto \ave{ \frac{1}{m} (E \omega^{z i_T}  - p^z \omega^{t i_T})\frac{p_{i_T}}{p_T} - \frac{p_T}{m} \omega^{tz}},
     \label{eq:RspinpT}
\end{align}
where the index $i_T$ sum over the transverse coordinates $i_T = x, y$. With a symmetric rapidity acceptance, the term proportional to $p^z$ should vanish. Then
\begin{equation}
    \Rspin(p_T) \propto \ave{\frac{E}{m}(\hat{p}^x \omega^{xz} + \hat{p}^y \omega^{yz}) - \frac{p_T}{m}\omega^{tz}},
    \label{eq:RspinpT_sym}
\end{equation}
where the unit 2D vector $\hat{p}^{i_T} \equiv p^{i_T}/p_T$ with $p_T = \sqrt{p_x^2 + p_y^2}$.
Eq.~\eqref{eq:RspinpT_sym} shows that $\Rspin(p_T)$ scales linearly with $p_T$ at large transverse momenta, which is observed in our calculations, shown in Fig.~\ref{fig:RspinpT_sys}. In the limit of $p_T \rightarrow 0$, $\Rspin(p_T) \rightarrow \ave{(\hat{p}^x \omega^{xz} + \hat{p}^y \omega^{yz})}$. Our calculations show $\Rspin(p_T) \rightarrow 0$ after the $\ave{\cdot}$ average defined in Eq.~\eqref{eq:thermav}.

For small systems like p+Au and $^3$He+Au collisions, the two sets of simulations ($f=0$ and $f=1$) give opposite slopes of $\Rspin(p_T)$ with $p_T$, which indicate the competition between the first and second terms in Eq.~\eqref{eq:RspinpT_sym}. The first term in Eq.~\eqref{eq:RspinpT_sym} is related to the vortex ring pattern, which is the reason we see a positive slope of $\Rspin(p_T)$ at high $p_T$ for the strong initial longitudinal flow case (with $f = 1$). For the other limit $f = 0$, the second term in Eq.~\eqref{eq:RspinpT_sym} dominates, which changes the slope of $\Rspin(p_T)$ to negative. Therefore, the measurement of $\Rspin(p_T)$ can provide information about the relative size between different components of the thermal vorticity tensor in the collision system.

Figure~\ref{fig:RspinpT_pAu_rap} further shows the $\Rspin(p_T)$ in three different rapidity windows for p+Au collisions. In the strong initial longitudinal flow case $f = 1$, the vortex ring pattern for the thermal vorticity stays roughly constant across rapidity. This flow pattern results in the almost rapidity independent $\Rspin(p_T)$ in panel (a). For the scenario with no initial longitudinal flow $(f = 0)$, the rapidity dependence of $\Rspin(p_T)$ is much more complex.

Our result reinforces the production plane polarization and the smoke ring variable as probes of longitudinal transparency and collective behaviors in small systems. 

\begin{figure}[h!]
  \centering
  \includegraphics[width=1.0\linewidth]{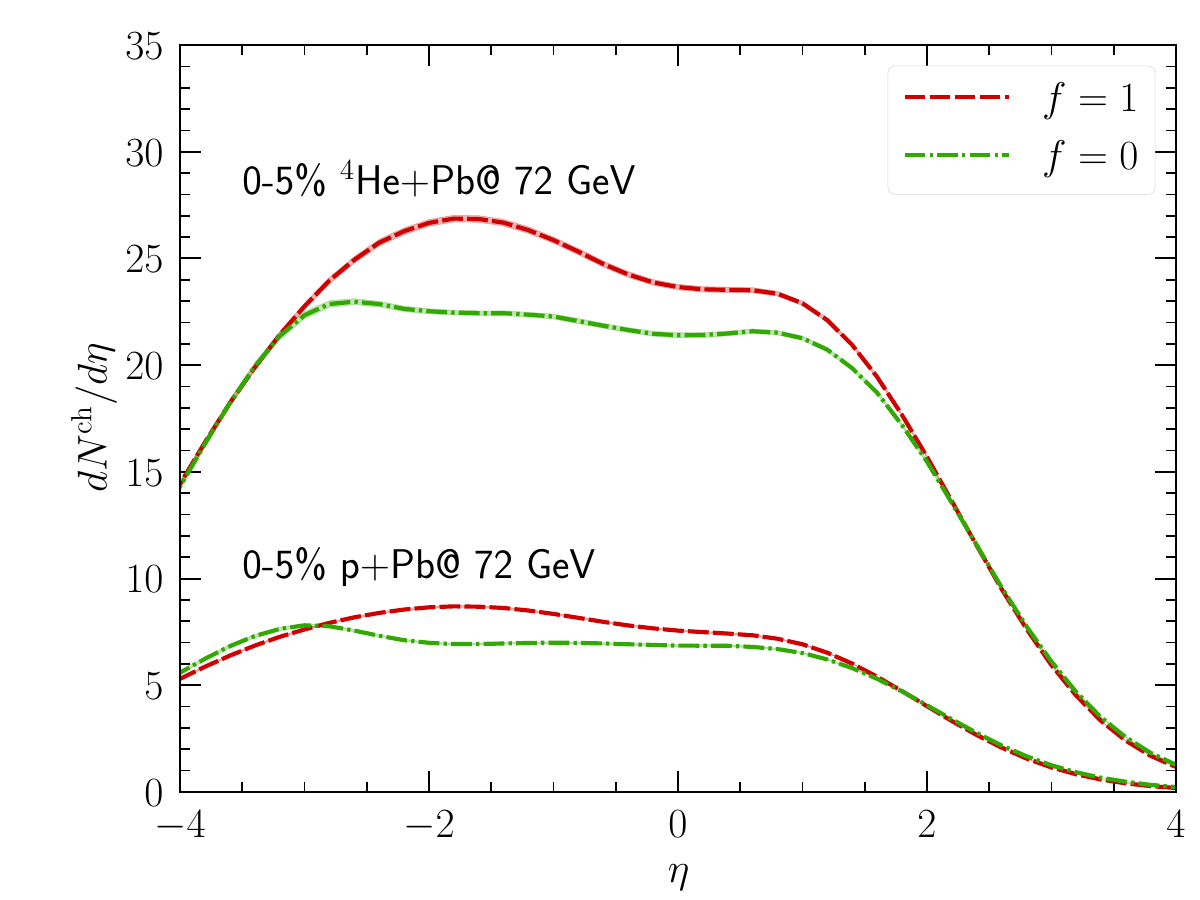}
  \caption{Charged hadron pseudorapidity distributions for 0-5\% p+Pb and $^4$He+Pb collisions at the fixed-target LHCb experiment. The negative rapidity denotes the Pb-going side.}
  \label{fig:dNdeta_72}
\end{figure}

\subsection{Vortex rings in the LHCb fixed-target SMOG experiment}

Now, we extend our simulations to asymmetric collision systems that are accessible in the upcoming LHCb fixed-target experiment with the System for Measuring Overlap with Gas (SMOG) setup.

Figure~\ref{fig:dNdeta_72} shows the charged hadron pseudo-rapidity distributions for 0-5\% central p+Pb and $^4$He+Pb collisions at $\snn = 72$~GeV. Our results show the same dependence as those at 200 GeV in Fig.~\ref{fig:dNdeta}.

\begin{figure}[h!]
  \centering
  \includegraphics[width=1.0\linewidth]{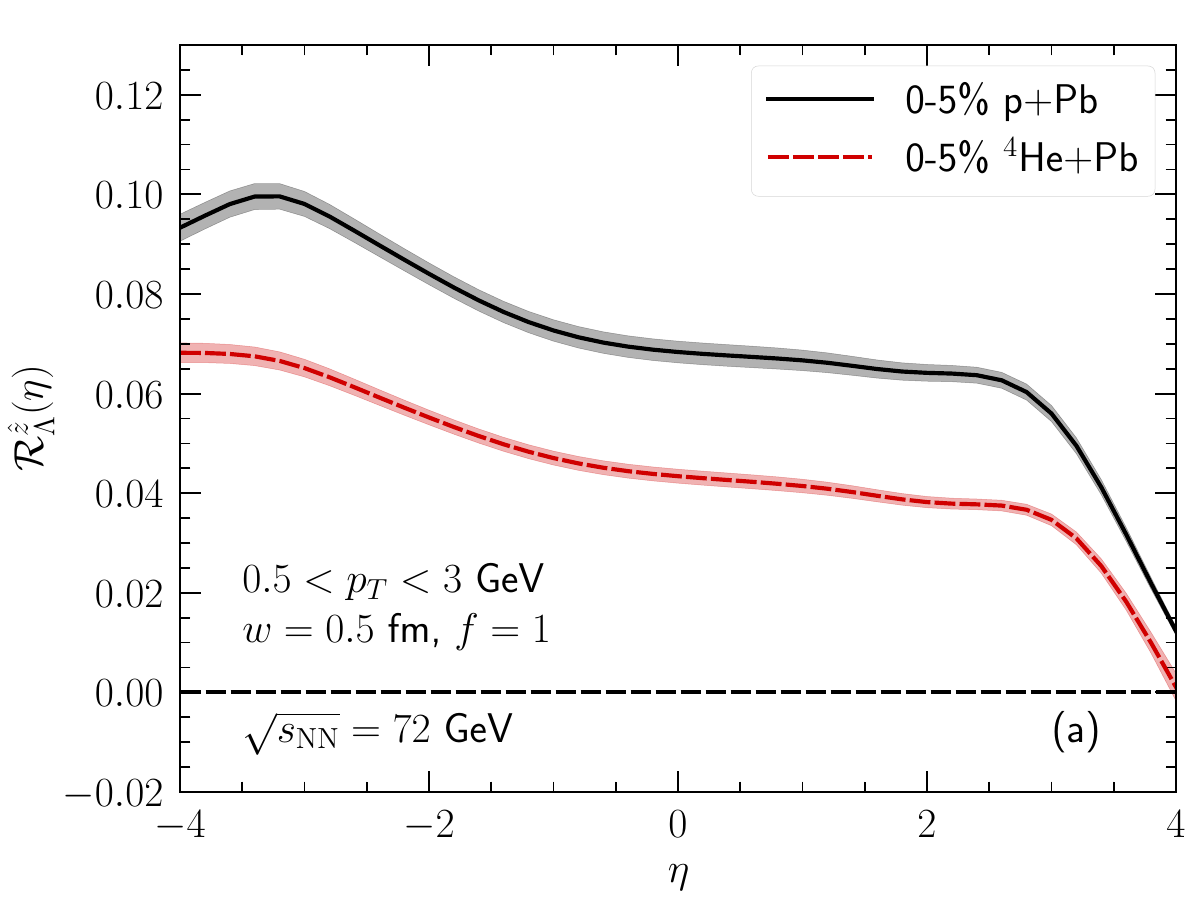}
  \includegraphics[width=1.0\linewidth]{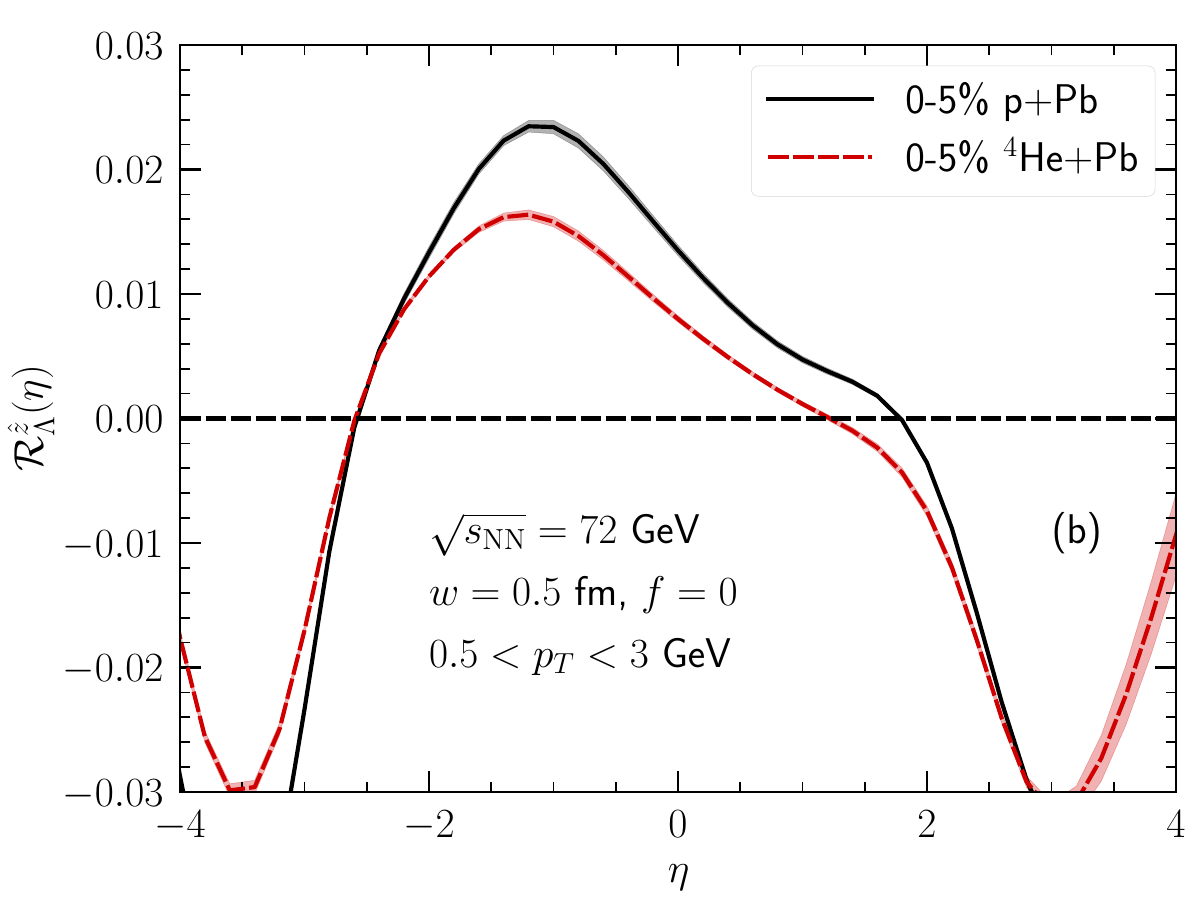}
  \caption{Model predictions for the hyperon's $\Rspin(\eta)$ in central p+Pb and $^4$He+Pb collisions at $\snn = 72$~GeV for the fixed-target LHCb SMOG experiment.}
  \label{fig:RspinEnergy}
\end{figure}

Figure~\ref{fig:RspinEnergy} shows the pseudo-rapidity dependence of the $\Rspin(\eta)$ in central p+Pb and $^4$He+Pb collisions at $\snn = 72$~GeV for two initial longitudinal flow scenarios. With strong initial longitudinal flow ($f = 1$), the small system results at $\snn = 72$~GeV agree qualitatively with the those at 200 GeV in Fig.~\ref{fig:RspinSys}a. The vortex ring structure extends over several units in rapidity in the Pb-going direction.
In the no initial longitudinal flow case ($f = 0$), Fig.~\ref{fig:RspinEnergy}b shows that the maximum $\Rspin(\eta)$ in central p+Pb and $^4$He+Pb collisions at $\snn = 72$~GeV could be about twice of the values in the similar small systems at 200 GeV, reflecting more non-trivial longitudinal dynamics in collision systems at the lower energy.

The LHCb SMOG experiment and the STAR forward spectrometer could perform such measurements at RHIC and LHC energies, as lower-energy fixed target runs would make the energy scan of this observable accessible.

\begin{figure}[t!]
  \centering
  \includegraphics[width=1.0\linewidth]{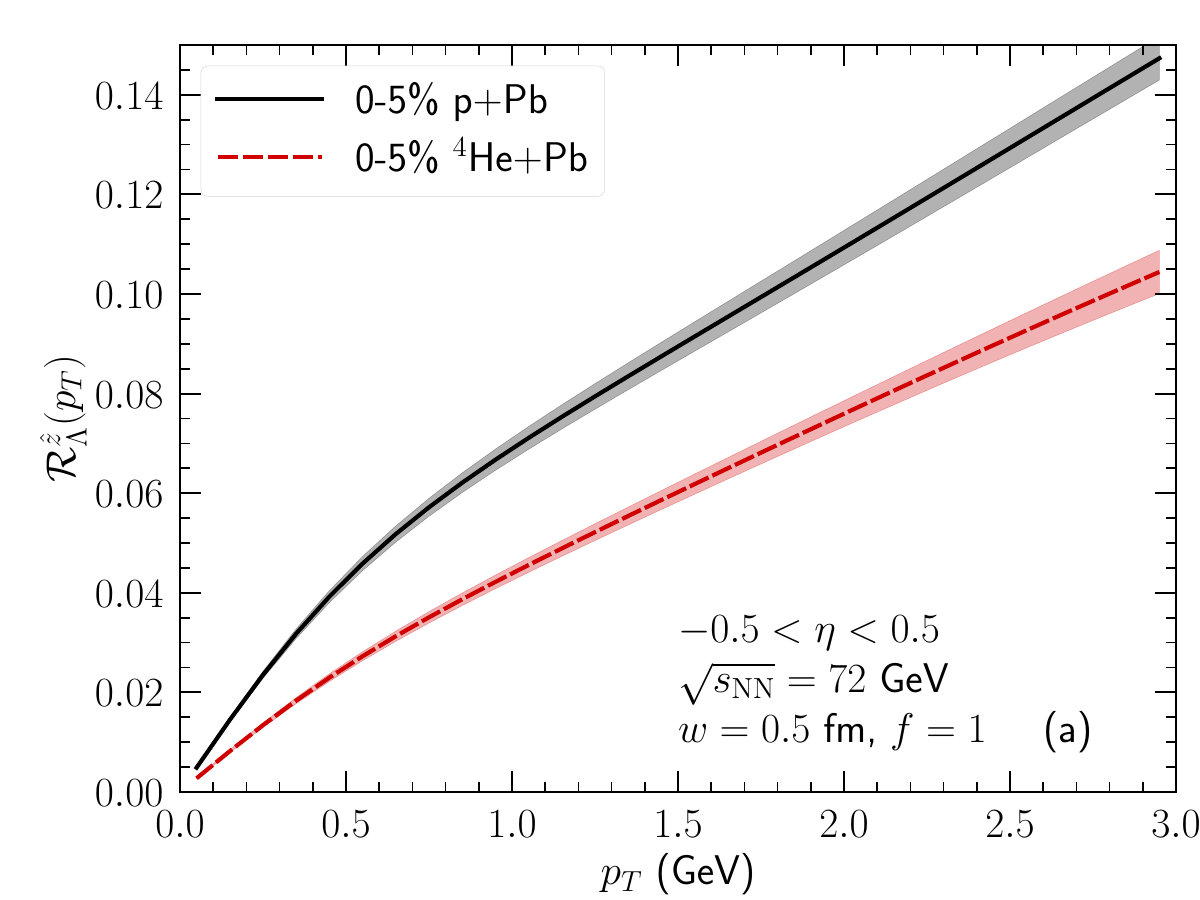}
  \includegraphics[width=1.0\linewidth]{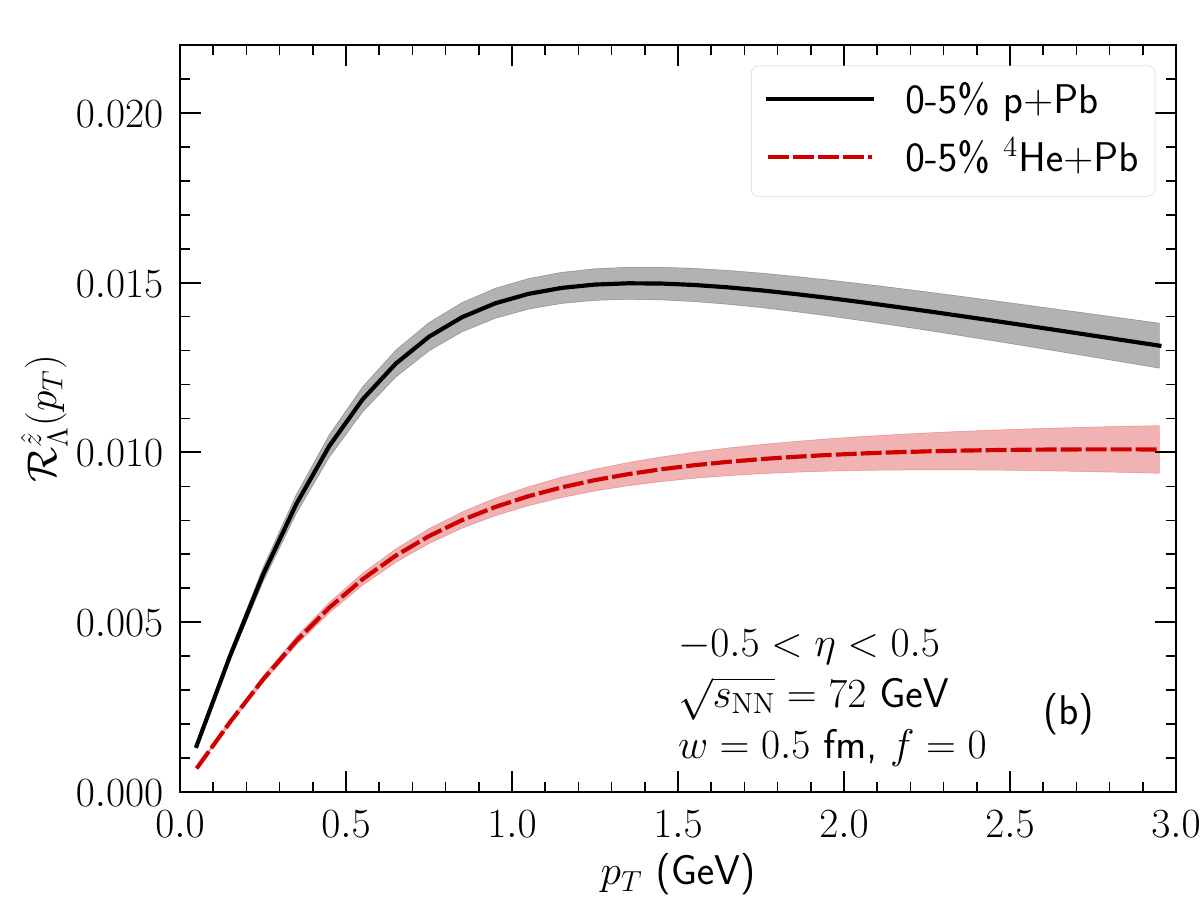}
  \caption{Model predictions for the hyperon's $\Rspin(\eta)$ in central p+Pb and $^4$He+Pb collisions at the fixed-target LHCb experiment.}
  \label{fig:Rspin_pT_LHCb_sys}
\end{figure}
\begin{figure}[t!]
  \centering
  \includegraphics[width=1.0\linewidth]{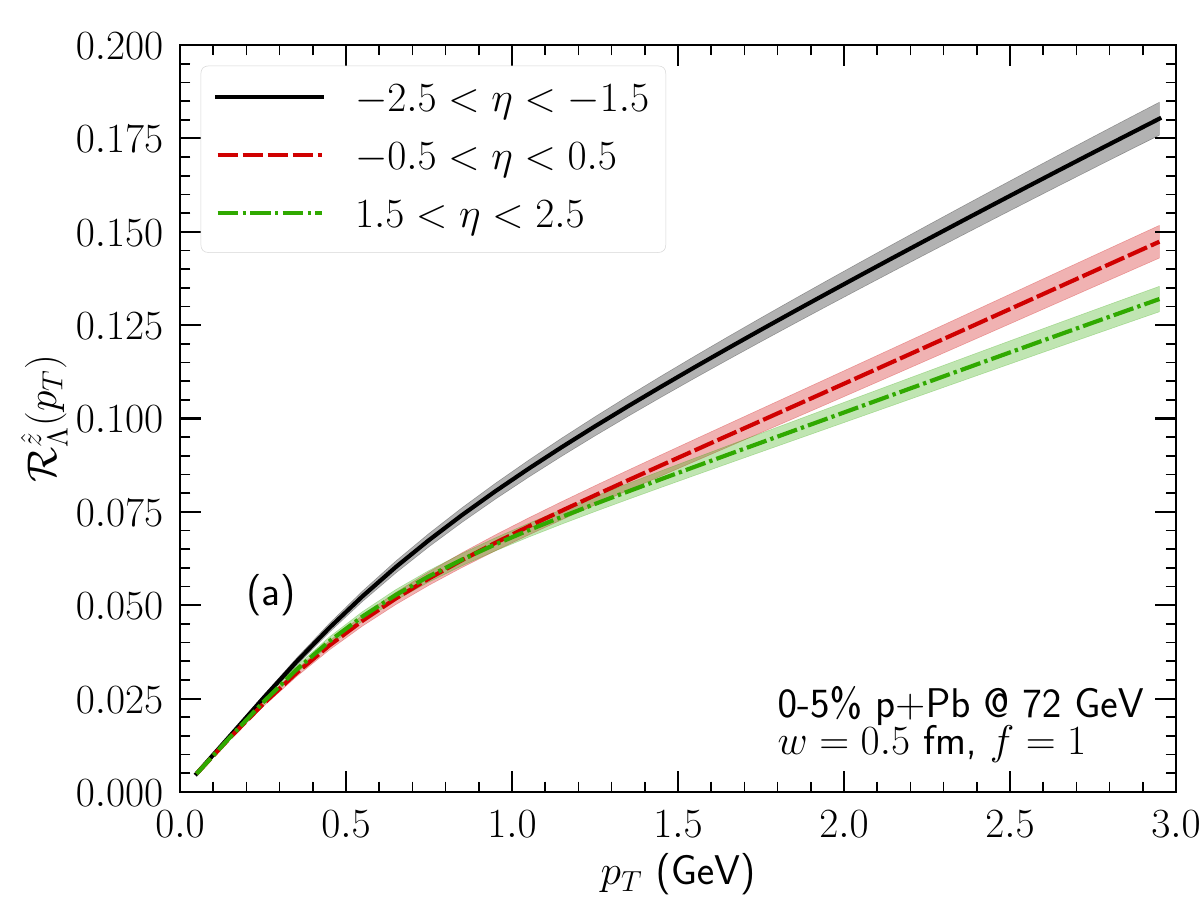}
  \includegraphics[width=1.0\linewidth]{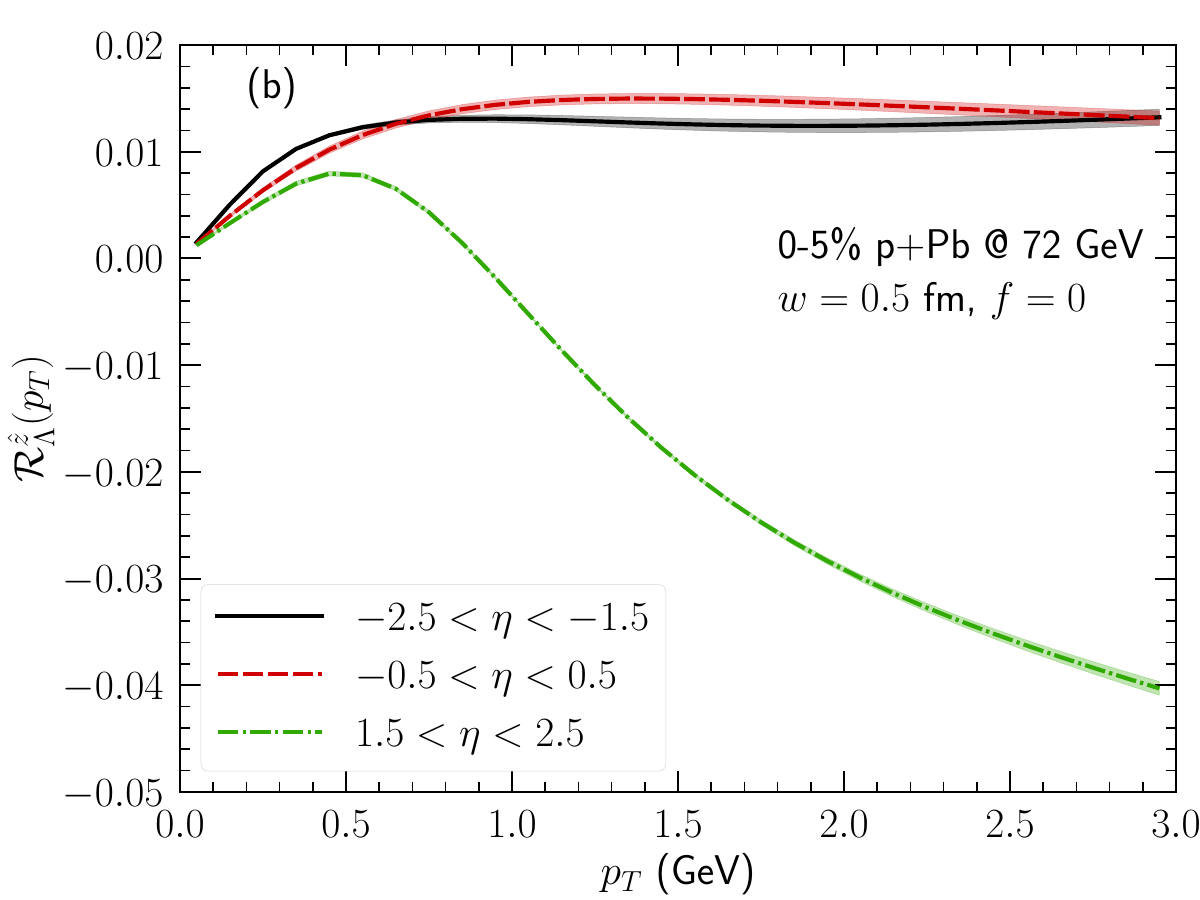}
  \caption{Model predictions for the hyperon's $\Rspin(\eta)$ in central p+Pb collisions at the fixed-target LHCb experiment.}
  \label{fig:Rspin_pT_LHCb_rap}
\end{figure}

Figures~\ref{fig:Rspin_pT_LHCb_sys} and \ref{fig:Rspin_pT_LHCb_rap} further show the $p_T$-differential $\Rspin(p_T)$ for small collision systems at the LHCb fixed-target collision energy. The results show a similar $p_T$ dependence at 200 GeV discussed above.

Overall, the potential $\Rspin$ measurements as functions of rapidity and transverse momentum at the LHCb fixed-target experiment would be a new observable for elucidating the collective nature of the produced collision systems.

\subsection{Model dependence of $\Rspin$}
\label{sec:modelDep}

To systematically explore the new $\Rspin$ observables, we introduce variations in the model parameters and identify how the proposed observables depend on these aspects.

\begin{figure}[ht!]
  \centering
  \includegraphics[width=0.96\linewidth]{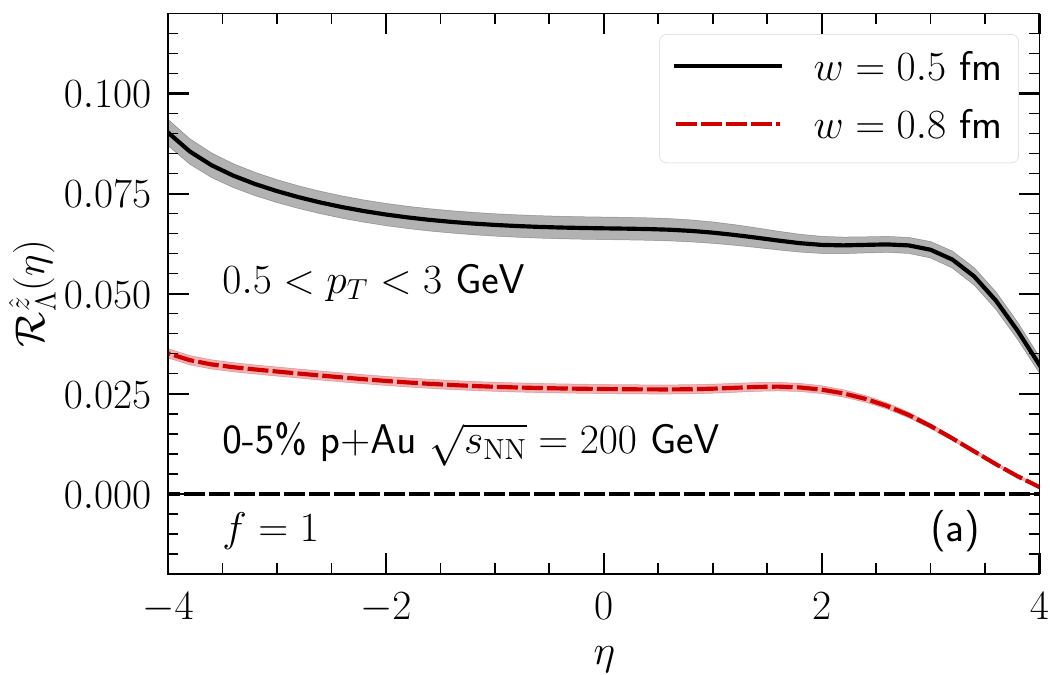}
  \includegraphics[width=0.96\linewidth]{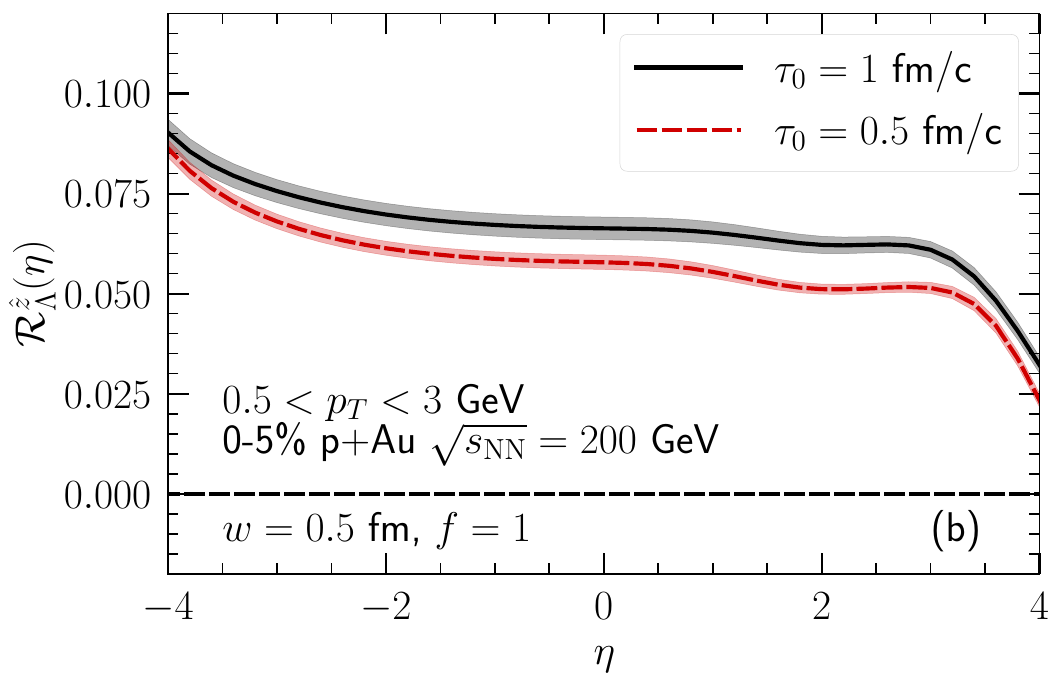}
  \includegraphics[width=0.96\linewidth]{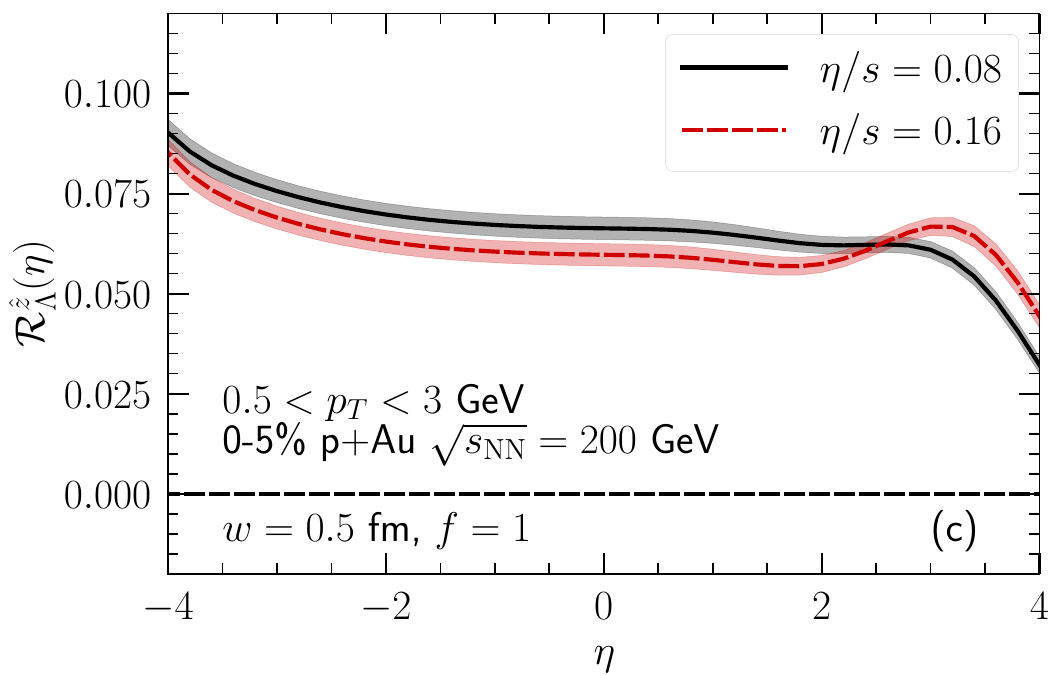}
  \includegraphics[width=0.96\linewidth]{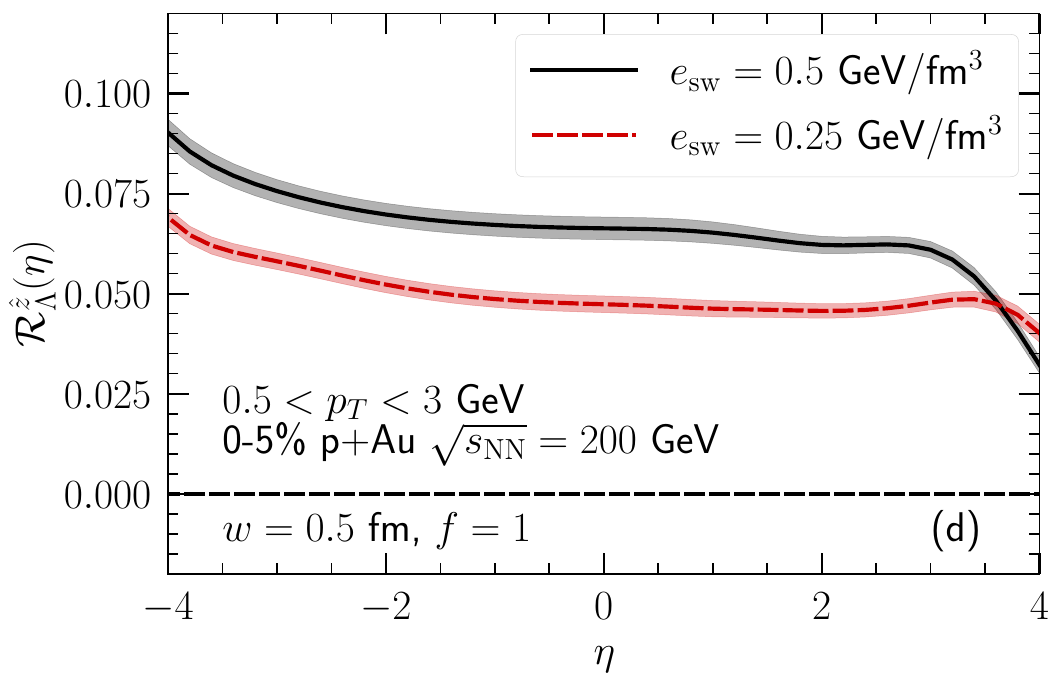}
  \caption{Model parameter dependence the hyperon's $\Rspin(\eta)$ in central p+Au collisions at $\snn = 200$ GeV.}
  \label{fig:Rspin_modelDep}
\end{figure}

\begin{figure}[ht!]
  \centering
  \includegraphics[width=0.96\linewidth]{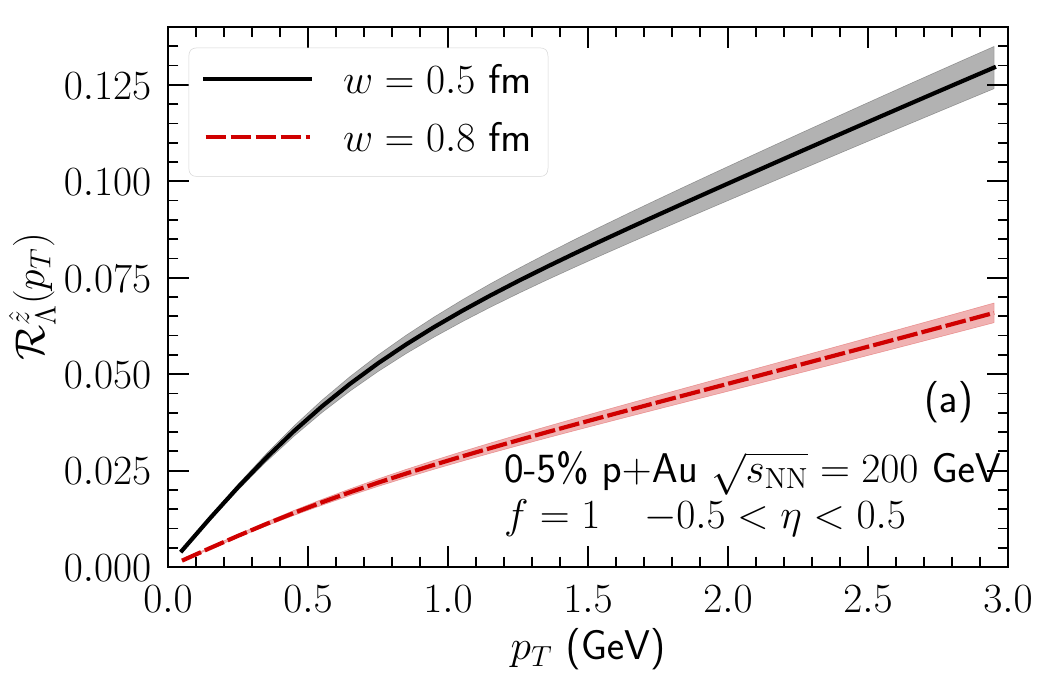}
  \includegraphics[width=0.96\linewidth]{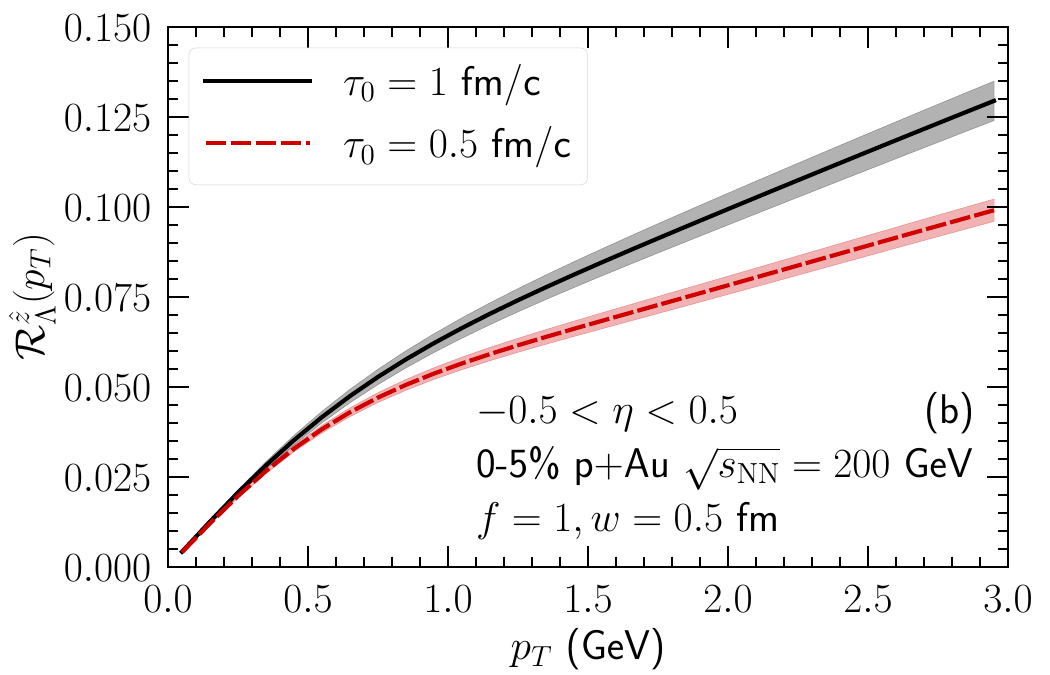}
  \includegraphics[width=0.96\linewidth]{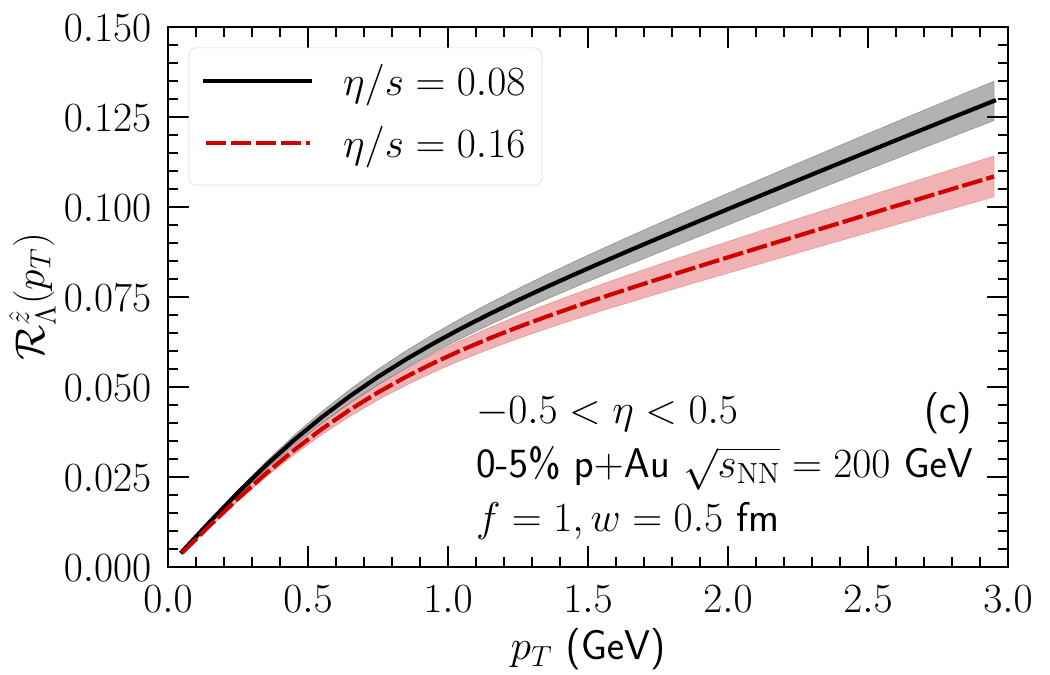}
  \includegraphics[width=0.96\linewidth]{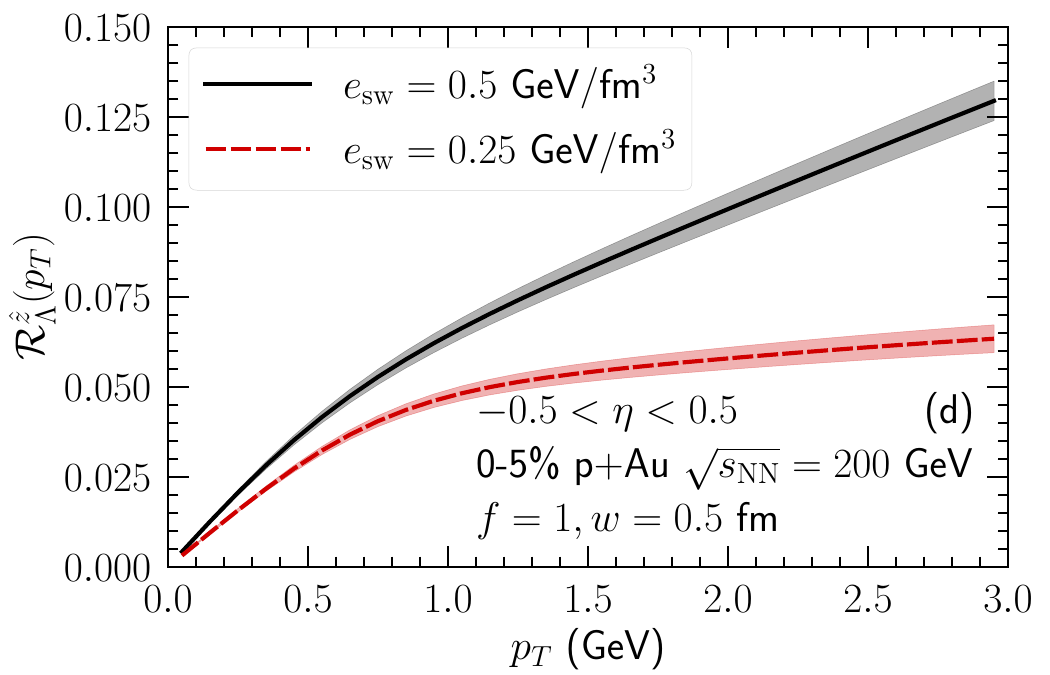}
  \caption{Model parameter dependence the hyperon's $\Rspin(p_T)$ in central p+Au collisions at $\snn = 200$ GeV.}
  \label{fig:Rspin_pT_modelDep}
\end{figure}

Figures~\ref{fig:Rspin_modelDep} and \ref{fig:Rspin_pT_modelDep} show how the $\Rspin(\eta)$ and  $\Rspin(p_T)$ vary with different model parameters in the simulations. We find that $\Rspin$ has a strong sensitivity on the initial hotspot's transverse size $w$. Small hotspots result in large temperature gradients in the early stages of the collisions and contribute to the thermal vorticity in the system. Although both a small hotspot size $w$ and a large $f$ in the model would result in large $\Rspin(\eta)$, the $p_T$-differential dependence in Fig.~\ref{fig:RspinpT_sys} can disentangle the two model parameter. Our prediction suggests a negative slope of $\Rspin(p_T)$ for no initial longitudinal flow case ($f = 0$).

Panels b, c, d in Figs.~\ref{fig:Rspin_modelDep} and \ref{fig:Rspin_pT_modelDep} show the model variations on the starting time of hydrodynamics $\tau_0$, QGP specific shear viscosity $\eta/s$, and freeze-out energy density $e_\mathrm{sw}$. We find that $\Rspin$ weakly depends on $\tau_0$ and $\eta/s$ in 0-5\% p+Au collisions at 200 GeV.

The value of $\Rspin$ is smaller with a lower freeze-out energy density $e_\mathrm{sw}$. A lower freeze-out energy density allows a longer fireball lifetime and more time for the system's thermal vorticity to relax to small values. The high $p_T$ slope of $\Rspin(p_T)$ shows a strong sensitivity to the value of $e_\mathrm{sw}$ used in the simulations.

\subsection{Additional gradient-induced polarization}

In addition to the thermal vorticity, the $\Lambda$ hyperons' polarization also receives contribution from the thermal shear tensor and gradients of $\mu_B/T$~\cite{Hidaka:2017auj, Liu:2020dxg, Liu:2021uhn, Becattini:2021suc, Buzzegoli:2021wlg, Liu:2021nyg},
\begin{align}
    S_\mathrm{SIP(type\,I)}^\mu(p) &= - \frac{1}{4m} \epsilon^{\mu\rho\sigma\tau} \ave{\frac{1}{p \cdot u} \hat{t}_\rho \xi_{\sigma \lambda} p^\lambda p_\tau}, \\
    S_\mathrm{SIP(type\,II)}^\mu(p) &= - \frac{1}{4m} \epsilon^{\mu\rho\sigma\tau} \ave{\frac{1}{p \cdot u} u_\rho \xi_{\sigma \lambda} p^\lambda_\perp p_\tau}, \\
    S_{\mu_B\mathrm{IP}}^\mu(p) &= - \frac{1}{4m} \epsilon^{\mu\rho\sigma\tau} \ave{\frac{T}{p \cdot u} u_\rho  \partial_\sigma \left(\frac{\mu_B}{T}\right) p_\tau}. \label{eq:muBIP}
\end{align}
Here, the thermal shear tensor $\xi^{\sigma \lambda} \equiv \frac{1}{2}[\partial^\sigma \left(\frac{u^\lambda}{T}\right) + \partial^\lambda \left(\frac{u^\sigma}{T}\right)]$. In the type I shear-induced polarization (SIP), the unit vector $\hat{t}^{\rho} = (1, 0, 0, 0)$~\cite{Becattini:2021suc, Buzzegoli:2021wlg}. In the type II SIP, $p^\lambda_\perp = p^\lambda - u^\lambda (p \cdot u)$ is the momentum vector transverse to the flow velocity~\cite{Liu:2020dxg, Liu:2021uhn}.

Using Eq.~\eqref{eq:RLambda3vectors}, we can obtain their contribution to the $\Rspin(p)$ observables as follows,
\begin{align}
    &\Rspin(\mathrm{SIP(type\,I)}) \propto \ave{\frac{p_\lambda}{m (p \cdot u)} [- \xi^{z \lambda} p_T -  \xi^{i_T \lambda} \hat{p}_{i_T}  p^z ]} \label{eq:RspinSIP1} \\
    &\Rspin(\mathrm{SIP(type\,II))} \propto \nonumber \\
    & \qquad \left\langle \frac{p_{\lambda, \perp}}{m (p \cdot u)}  \{- \xi^{z \lambda} [p^t (u^{i_T} \hat{p}_{i_T}) + u^t p_T] \right. \nonumber \\
    & \hspace{2.4cm} - \xi^{i_T \lambda} \hat{p}_{i_T} (u^t p^z  - u^z p^t) \nonumber \\
    & \hspace{2.4cm} + \xi^{t \lambda} [(u^{i_T} \hat{p}_{i_T}) p^z + u^z p_T] \} \bigg\rangle  \label{eq:RspinSIP2} \\
    &\Rspin(\mu_B\mathrm{IP}) \propto \nonumber \\
    & \qquad \bigg\langle \frac{T}{m (p \cdot u)} \bigg\{ -\partial^z \left(\frac{\mu_B}{T}\right) [p^t (u^{i_T} \hat{p}_{i_T}) + u^t p_T] \nonumber \\
    & \hspace{2.4cm} - \hat{p}_{i_T} \partial^{i_T} \left(\frac{\mu_B}{T}\right) (u^t p^z - p^t u^z) \nonumber \\
    & \hspace{2.4cm} + \partial^t \left(\frac{\mu_B}{T}\right) [(u^{i_T} \hat{p}_{i_T}) p^z + u^z p_T] \bigg\} \bigg\rangle  \label{eq:RspinmuBIP}
\end{align}
Here, the index $i_T$ runs over the transverse coordinates, $i_T = x, y$ and $\hat{p}^{i_T} = p^{i_T}/p_T$ is the unit 2D vector for particle's transverse momentum.

\begin{figure}[ht!]
  \centering
  \includegraphics[width=1.0\linewidth]{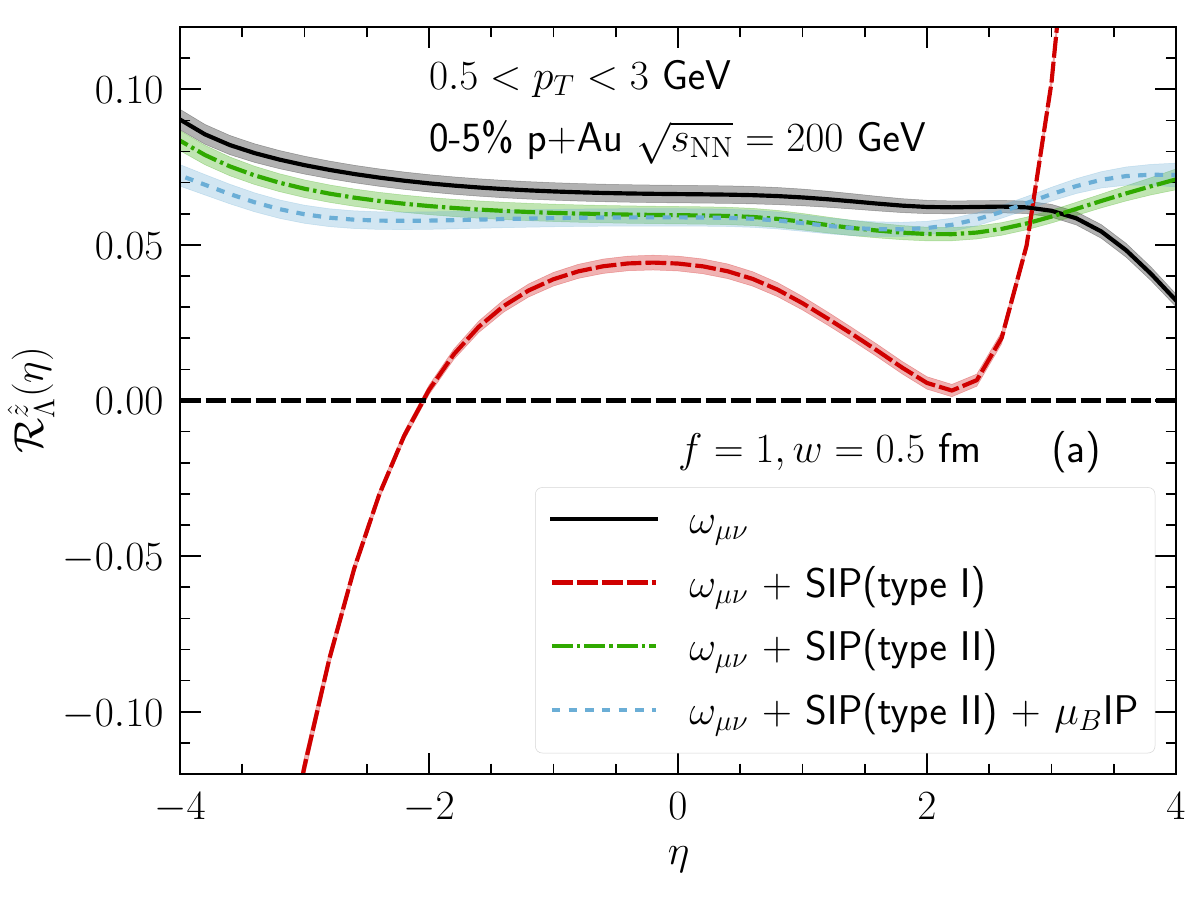}
  \includegraphics[width=1.0\linewidth]{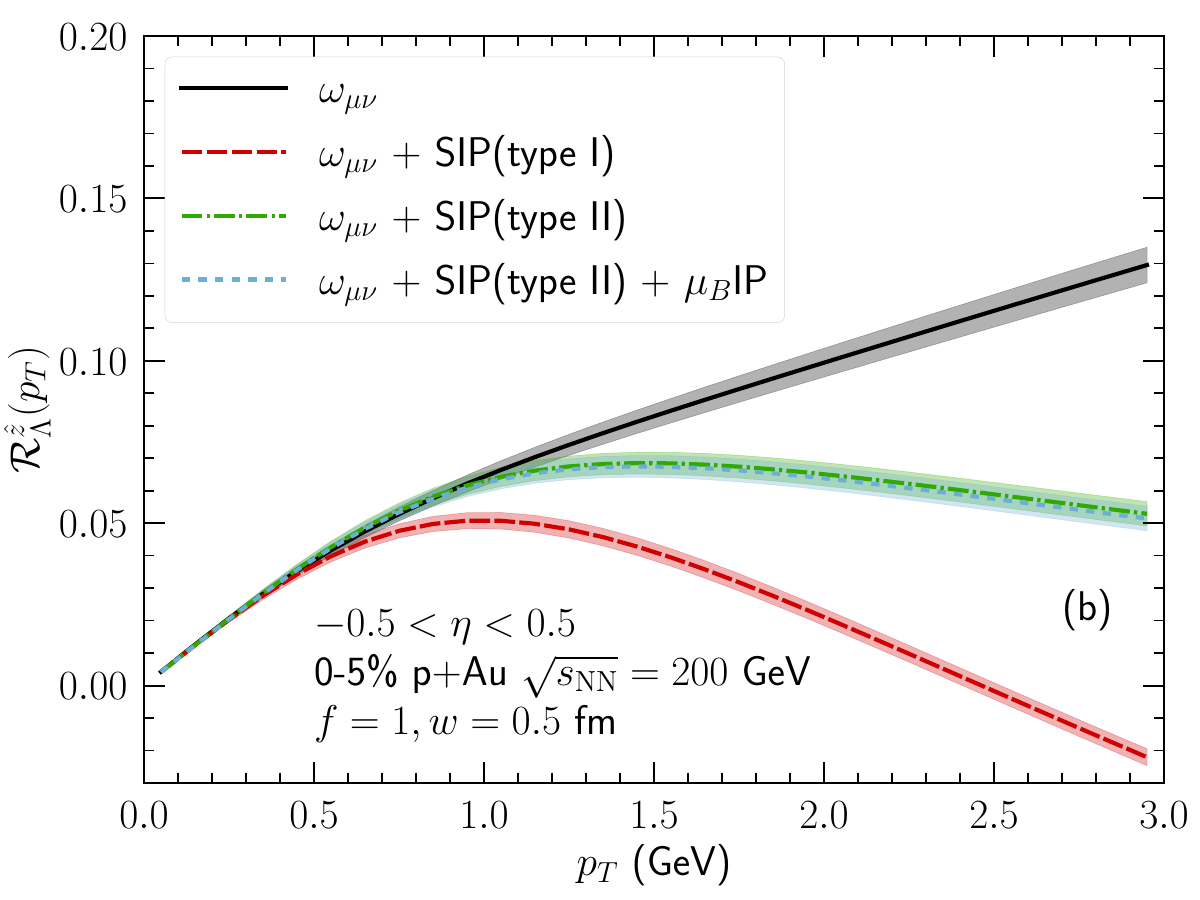}
  \caption{The pseudorapidity (a) and $p_T$ (b) dependencies of $\Rspin$ observables with additional polarization induced by the gradients of thermal shear tensor and baryon chemical potential.}
  \label{fig:symmetricShear}
\end{figure}

Figure~\ref{fig:symmetricShear} shows the numerical results of these gradient-induced polarization contributions to the ring observable $\Rspin$. We observe that the shear-induced polarization (type I) has a large contribution to $\Rspin$, especially at large rapidity regions. While the other SIP formulation (type II) gives small corrections to the $\Rspin$ on top of that from the thermal vorticity.
Our result indicates a substantial theoretical uncertainty from the SIP contribution to the observable vortex ring.
This result demonstrates that measurements of the vortex ring observable can differentiate the form of symmetric shear contribution to $\Lambda$'s polarization.
Figure~\ref{fig:symmetricShear}b shows that the contributions from shear-induced polarization to the $\Rspin$ observable grow with $\Lambda$'s transverse momentum. Eqs.~\eqref{eq:RspinSIP1} and \eqref{eq:RspinSIP2} indicate that the SIP contribution scales linearly with $p_T$, which is seen in the numerical results in Fig.~\ref{fig:symmetricShear}b. Finally, we find that the contribution from the gradients of $\mu_B/T$ in Eq.~\eqref{eq:muBIP} do not play a significant role in small systems at 200 GeV.

\section{Conclusions} \label{sec:conc}

In conclusion, we have confirmed and extended the studies of the production plane polarization observable $\Rspin$ in asymmetric collisions, first discussed in \cite{Lisa:2021zkj}.  We included hot spot and impact parameter fluctuations event-by-event and expanded our scope to study its system size dependence. The pseudo-rapidity dependence of the $\Rspin(\eta)$ observable can serve as a sensitive probe for the initial longitudinal flow velocity. We predict a linear $p_T$ dependence of $\Rspin(p_T)$ for $p_T > 1$ GeV at mid-rapidity. Its slope can provide information about the relative sizes of individual components of the thermal vorticity tensor. 
We further provide model predictions for asymmetric collision systems in the LHCb SMOG experiment setup. These asymmetric collisions at $\snn = 72$~GeV reveal more non-trivial longitudinal dynamics. 
We systematically explored the sensitivity of the $\Rspin$ observable on various model parameters. 

The proposed $\Rspin$ observable is a promising probe of hydrodynamic behavior in small asymmetric systems. Quantitative comparisons will set valuable constraints on the longitudinal dynamics and early-stage stopping mechanism.
We look forward to experimental investigations in this direction.

\section*{Acknowledgments}
This work is in part supported by the U.S. Department of Energy (DOE) under award numbers DE-SC0021969 and DE-SC0020651.
C.S. acknowledges a DOE Office of Science Early Career Award.
M.L. acknowledges the support of the Fulbright Commission of Brazil.
J.T.~was supported by FAPESP projects 2017/05685-2 and CNPq through 309174/2020-1.
G.T.~acknowledges support from Bolsa de produtividade CNPQ 305731/2023-8, Bolsa de pesquisa FAPESP 2023/06278-2.
This research was done using resources provided by the Open Science Grid (OSG)~\cite{Pordes:2007zzb, Sfiligoi:2009cct, https://doi.org/10.21231/906p-4d78, https://doi.org/10.21231/0kvz-ve57}, which is supported by the National Science Foundation award \#2030508 and \#1836650.

\bibliography{ebeVorticityRing, noninspires}

\begin{thebibliography}{39}%
\makeatletter
\providecommand \@ifxundefined [1]{%
 \@ifx{#1\undefined}
}%
\providecommand \@ifnum [1]{%
 \ifnum #1\expandafter \@firstoftwo
 \else \expandafter \@secondoftwo
 \fi
}%
\providecommand \@ifx [1]{%
 \ifx #1\expandafter \@firstoftwo
 \else \expandafter \@secondoftwo
 \fi
}%
\providecommand \natexlab [1]{#1}%
\providecommand \enquote  [1]{``#1''}%
\providecommand \bibnamefont  [1]{#1}%
\providecommand \bibfnamefont [1]{#1}%
\providecommand \citenamefont [1]{#1}%
\providecommand \href@noop [0]{\@secondoftwo}%
\providecommand \href [0]{\begingroup \@sanitize@url \@href}%
\providecommand \@href[1]{\@@startlink{#1}\@@href}%
\providecommand \@@href[1]{\endgroup#1\@@endlink}%
\providecommand \@sanitize@url [0]{\catcode `\\12\catcode `\$12\catcode
  `\&12\catcode `\#12\catcode `\^12\catcode `\_12\catcode `\%12\relax}%
\providecommand \@@startlink[1]{}%
\providecommand \@@endlink[0]{}%
\providecommand \url  [0]{\begingroup\@sanitize@url \@url }%
\providecommand \@url [1]{\endgroup\@href {#1}{\urlprefix }}%
\providecommand \urlprefix  [0]{URL }%
\providecommand \Eprint [0]{\href }%
\providecommand \doibase [0]{http://dx.doi.org/}%
\providecommand \selectlanguage [0]{\@gobble}%
\providecommand \bibinfo  [0]{\@secondoftwo}%
\providecommand \bibfield  [0]{\@secondoftwo}%
\providecommand \translation [1]{[#1]}%
\providecommand \BibitemOpen [0]{}%
\providecommand \bibitemStop [0]{}%
\providecommand \bibitemNoStop [0]{.\EOS\space}%
\providecommand \EOS [0]{\spacefactor3000\relax}%
\providecommand \BibitemShut  [1]{\csname bibitem#1\endcsname}%
\let\auto@bib@innerbib\@empty
\bibitem [{\citenamefont {Helmholtz}(1867)}]{doi:10.1080/14786446708639824}%
  \BibitemOpen
  \bibfield  {author} {\bibinfo {author} {\bibfnamefont {H.}~\bibnamefont
  {Helmholtz}},\ }\bibfield  {title} {\enquote {\bibinfo {title} {Lxiii. on
  integrals of the hydrodynamical equations, which express vortex-motion},}\
  }\href {\doibase 10.1080/14786446708639824} {\bibfield  {journal} {\bibinfo
  {journal} {The London, Edinburgh, and Dublin Philosophical Magazine and
  Journal of Science}\ }\textbf {\bibinfo {volume} {33}},\ \bibinfo {pages}
  {485--512} (\bibinfo {year} {1867})},\ \Eprint
  {http://arxiv.org/abs/https://doi.org/10.1080/14786446708639824}
  {https://doi.org/10.1080/14786446708639824} \BibitemShut {NoStop}%
\bibitem [{\citenamefont {Dusling}\ \emph {et~al.}(2016)\citenamefont
  {Dusling}, \citenamefont {Li},\ and\ \citenamefont
  {Schenke}}]{Dusling:2015gta}%
  \BibitemOpen
  \bibfield  {author} {\bibinfo {author} {\bibfnamefont {Kevin}\ \bibnamefont
  {Dusling}}, \bibinfo {author} {\bibfnamefont {Wei}\ \bibnamefont {Li}}, \
  and\ \bibinfo {author} {\bibfnamefont {Bj\"orn}\ \bibnamefont {Schenke}},\
  }\bibfield  {title} {\enquote {\bibinfo {title} {{Novel collective phenomena
  in high-energy proton\textendash{}proton and proton\textendash{}nucleus
  collisions}},}\ }\href {\doibase 10.1142/S0218301316300022} {\bibfield
  {journal} {\bibinfo  {journal} {Int. J. Mod. Phys. E}\ }\textbf {\bibinfo
  {volume} {25}},\ \bibinfo {pages} {1630002} (\bibinfo {year} {2016})},\
  \Eprint {http://arxiv.org/abs/1509.07939} {arXiv:1509.07939 [nucl-ex]}
  \BibitemShut {NoStop}%
\bibitem [{\citenamefont {Nagle}\ and\ \citenamefont
  {Zajc}(2018)}]{Nagle:2018nvi}%
  \BibitemOpen
  \bibfield  {author} {\bibinfo {author} {\bibfnamefont {James~L.}\
  \bibnamefont {Nagle}}\ and\ \bibinfo {author} {\bibfnamefont {William~A.}\
  \bibnamefont {Zajc}},\ }\bibfield  {title} {\enquote {\bibinfo {title}
  {{Small System Collectivity in Relativistic Hadronic and Nuclear
  Collisions}},}\ }\href {\doibase 10.1146/annurev-nucl-101916-123209}
  {\bibfield  {journal} {\bibinfo  {journal} {Ann. Rev. Nucl. Part. Sci.}\
  }\textbf {\bibinfo {volume} {68}},\ \bibinfo {pages} {211--235} (\bibinfo
  {year} {2018})},\ \Eprint {http://arxiv.org/abs/1801.03477} {arXiv:1801.03477
  [nucl-ex]} \BibitemShut {NoStop}%
\bibitem [{\citenamefont {Shen}\ and\ \citenamefont
  {Yan}(2020)}]{Shen:2020mgh}%
  \BibitemOpen
  \bibfield  {author} {\bibinfo {author} {\bibfnamefont {Chun}\ \bibnamefont
  {Shen}}\ and\ \bibinfo {author} {\bibfnamefont {Li}~\bibnamefont {Yan}},\
  }\bibfield  {title} {\enquote {\bibinfo {title} {{Recent development of
  hydrodynamic modeling in heavy-ion collisions}},}\ }\href {\doibase
  10.1007/s41365-020-00829-z} {\bibfield  {journal} {\bibinfo  {journal} {Nucl.
  Sci. Tech.}\ }\textbf {\bibinfo {volume} {31}},\ \bibinfo {pages} {122}
  (\bibinfo {year} {2020})},\ \Eprint {http://arxiv.org/abs/2010.12377}
  {arXiv:2010.12377 [nucl-th]} \BibitemShut {NoStop}%
\bibitem [{\citenamefont {Schenke}(2021)}]{Schenke:2021mxx}%
  \BibitemOpen
  \bibfield  {author} {\bibinfo {author} {\bibfnamefont {Bj\"orn}\ \bibnamefont
  {Schenke}},\ }\bibfield  {title} {\enquote {\bibinfo {title} {{The smallest
  fluid on Earth}},}\ }\href {\doibase 10.1088/1361-6633/ac14c9} {\bibfield
  {journal} {\bibinfo  {journal} {Rept. Prog. Phys.}\ }\textbf {\bibinfo
  {volume} {84}},\ \bibinfo {pages} {082301} (\bibinfo {year} {2021})},\
  \Eprint {http://arxiv.org/abs/2102.11189} {arXiv:2102.11189 [nucl-th]}
  \BibitemShut {NoStop}%
\bibitem [{\citenamefont {Noronha}\ \emph {et~al.}(2024)\citenamefont
  {Noronha}, \citenamefont {Schenke}, \citenamefont {Shen},\ and\ \citenamefont
  {Zhao}}]{Noronha:2024dtq}%
  \BibitemOpen
  \bibfield  {author} {\bibinfo {author} {\bibfnamefont {Jorge}\ \bibnamefont
  {Noronha}}, \bibinfo {author} {\bibfnamefont {Bj\"orn}\ \bibnamefont
  {Schenke}}, \bibinfo {author} {\bibfnamefont {Chun}\ \bibnamefont {Shen}}, \
  and\ \bibinfo {author} {\bibfnamefont {Wenbin}\ \bibnamefont {Zhao}},\
  }\bibfield  {title} {\enquote {\bibinfo {title} {{Progress and Challenges in
  Small Systems}},}\ \ }(\bibinfo {year} {2024})\ \Eprint
  {http://arxiv.org/abs/2401.09208} {arXiv:2401.09208 [nucl-th]} \BibitemShut
  {NoStop}%
\bibitem [{\citenamefont {Lisa}\ \emph {et~al.}(2021)\citenamefont {Lisa},
  \citenamefont {Barbon}, \citenamefont {Chinellato}, \citenamefont {Serenone},
  \citenamefont {Shen}, \citenamefont {Takahashi},\ and\ \citenamefont
  {Torrieri}}]{Lisa:2021zkj}%
  \BibitemOpen
  \bibfield  {author} {\bibinfo {author} {\bibfnamefont {Michael~Annan}\
  \bibnamefont {Lisa}}, \bibinfo {author} {\bibfnamefont {Jo\~ao
  Guilherme~Prado}\ \bibnamefont {Barbon}}, \bibinfo {author} {\bibfnamefont
  {David~Dobrigkeit}\ \bibnamefont {Chinellato}}, \bibinfo {author}
  {\bibfnamefont {Willian~Matioli}\ \bibnamefont {Serenone}}, \bibinfo {author}
  {\bibfnamefont {Chun}\ \bibnamefont {Shen}}, \bibinfo {author} {\bibfnamefont
  {Jun}\ \bibnamefont {Takahashi}}, \ and\ \bibinfo {author} {\bibfnamefont
  {Giorgio}\ \bibnamefont {Torrieri}},\ }\bibfield  {title} {\enquote {\bibinfo
  {title} {{Vortex rings from high energy central p+A collisions}},}\ }\href
  {\doibase 10.1103/PhysRevC.104.L011901} {\bibfield  {journal} {\bibinfo
  {journal} {Phys. Rev. C}\ }\textbf {\bibinfo {volume} {104}},\ \bibinfo
  {pages} {011901} (\bibinfo {year} {2021})},\ \Eprint
  {http://arxiv.org/abs/2101.10872} {arXiv:2101.10872 [hep-ph]} \BibitemShut
  {NoStop}%
\bibitem [{\citenamefont {Serenone}\ \emph {et~al.}(2021)\citenamefont
  {Serenone}, \citenamefont {Barbon}, \citenamefont {Chinellato}, \citenamefont
  {Lisa}, \citenamefont {Shen}, \citenamefont {Takahashi},\ and\ \citenamefont
  {Torrieri}}]{jet1}%
  \BibitemOpen
  \bibfield  {author} {\bibinfo {author} {\bibfnamefont {Willian~Matioli}\
  \bibnamefont {Serenone}}, \bibinfo {author} {\bibfnamefont {Jo\~ao
  Guilherme~Prado}\ \bibnamefont {Barbon}}, \bibinfo {author} {\bibfnamefont
  {David~Dobrigkeit}\ \bibnamefont {Chinellato}}, \bibinfo {author}
  {\bibfnamefont {Michael~Annan}\ \bibnamefont {Lisa}}, \bibinfo {author}
  {\bibfnamefont {Chun}\ \bibnamefont {Shen}}, \bibinfo {author} {\bibfnamefont
  {Jun}\ \bibnamefont {Takahashi}}, \ and\ \bibinfo {author} {\bibfnamefont
  {Giorgio}\ \bibnamefont {Torrieri}},\ }\bibfield  {title} {\enquote {\bibinfo
  {title} {{\ensuremath{\Lambda} polarization from thermalized jet energy}},}\
  }\href {\doibase 10.1016/j.physletb.2021.136500} {\bibfield  {journal}
  {\bibinfo  {journal} {Phys. Lett. B}\ }\textbf {\bibinfo {volume} {820}},\
  \bibinfo {pages} {136500} (\bibinfo {year} {2021})},\ \Eprint
  {http://arxiv.org/abs/2102.11919} {arXiv:2102.11919 [hep-ph]} \BibitemShut
  {NoStop}%
\bibitem [{\citenamefont {Ribeiro}\ \emph {et~al.}(2024)\citenamefont
  {Ribeiro}, \citenamefont {Dobrigkeit~Chinellato}, \citenamefont {Lisa},
  \citenamefont {Matioli~Serenone}, \citenamefont {Shen}, \citenamefont
  {Takahashi},\ and\ \citenamefont {Torrieri}}]{jet2}%
  \BibitemOpen
  \bibfield  {author} {\bibinfo {author} {\bibfnamefont {Vitor~Hugo}\
  \bibnamefont {Ribeiro}}, \bibinfo {author} {\bibfnamefont {David}\
  \bibnamefont {Dobrigkeit~Chinellato}}, \bibinfo {author} {\bibfnamefont
  {Michael~Annan}\ \bibnamefont {Lisa}}, \bibinfo {author} {\bibfnamefont
  {Willian}\ \bibnamefont {Matioli~Serenone}}, \bibinfo {author} {\bibfnamefont
  {Chun}\ \bibnamefont {Shen}}, \bibinfo {author} {\bibfnamefont {Jun}\
  \bibnamefont {Takahashi}}, \ and\ \bibinfo {author} {\bibfnamefont {Giorgio}\
  \bibnamefont {Torrieri}},\ }\bibfield  {title} {\enquote {\bibinfo {title}
  {{\ensuremath{\Lambda} polarization from vortex rings as the medium response
  for jet thermalization}},}\ }\href {\doibase 10.1103/PhysRevC.109.014905}
  {\bibfield  {journal} {\bibinfo  {journal} {Phys. Rev. C}\ }\textbf {\bibinfo
  {volume} {109}},\ \bibinfo {pages} {014905} (\bibinfo {year} {2024})},\
  \Eprint {http://arxiv.org/abs/2305.02428} {arXiv:2305.02428 [hep-ph]}
  \BibitemShut {NoStop}%
\bibitem [{\citenamefont {Heller}\ \emph {et~al.}(1977)\citenamefont {Heller},
  \citenamefont {Overseth}, \citenamefont {Bunce}, \citenamefont {Dydak},\ and\
  \citenamefont {Taureg}}]{Heller:1977mv}%
  \BibitemOpen
  \bibfield  {author} {\bibinfo {author} {\bibfnamefont {Kenneth~J.}\
  \bibnamefont {Heller}}, \bibinfo {author} {\bibfnamefont {O.~E.}\
  \bibnamefont {Overseth}}, \bibinfo {author} {\bibfnamefont {G.}~\bibnamefont
  {Bunce}}, \bibinfo {author} {\bibfnamefont {F.}~\bibnamefont {Dydak}}, \ and\
  \bibinfo {author} {\bibfnamefont {H.}~\bibnamefont {Taureg}},\ }\bibfield
  {title} {\enquote {\bibinfo {title} {{Lambda0 Hyperon Polarization in
  Inclusive Production by 24-GeV Protons on Platinum}},}\ }\href {\doibase
  10.1016/0370-2693(77)90476-2} {\bibfield  {journal} {\bibinfo  {journal}
  {Phys. Lett. B}\ }\textbf {\bibinfo {volume} {68}},\ \bibinfo {pages}
  {480--482} (\bibinfo {year} {1977})}\BibitemShut {NoStop}%
\bibitem [{\citenamefont {Hauenstein}\ \emph {et~al.}(2016)\citenamefont
  {Hauenstein} \emph {et~al.}}]{Hauenstein:2016som}%
  \BibitemOpen
  \bibfield  {author} {\bibinfo {author} {\bibfnamefont {F.}~\bibnamefont
  {Hauenstein}} \emph {et~al.} (\bibinfo {collaboration} {COSY-TOF}),\
  }\bibfield  {title} {\enquote {\bibinfo {title} {{Measurement of polarization
  observables of the associated strangeness production in proton proton
  interactions}},}\ }\href {\doibase 10.1140/epja/i2016-16337-1} {\bibfield
  {journal} {\bibinfo  {journal} {Eur. Phys. J. A}\ }\textbf {\bibinfo {volume}
  {52}},\ \bibinfo {pages} {337} (\bibinfo {year} {2016})},\ \Eprint
  {http://arxiv.org/abs/1607.06305} {arXiv:1607.06305 [nucl-ex]} \BibitemShut
  {NoStop}%
\bibitem [{\citenamefont {Bunce}\ \emph {et~al.}(1976)\citenamefont {Bunce}
  \emph {et~al.}}]{Bunce:1976yb}%
  \BibitemOpen
  \bibfield  {author} {\bibinfo {author} {\bibfnamefont {G.}~\bibnamefont
  {Bunce}} \emph {et~al.},\ }\bibfield  {title} {\enquote {\bibinfo {title}
  {{Lambda0 Hyperon Polarization in Inclusive Production by 300-GeV Protons on
  Beryllium.}}}\ }\href {\doibase 10.1103/PhysRevLett.36.1113} {\bibfield
  {journal} {\bibinfo  {journal} {Phys. Rev. Lett.}\ }\textbf {\bibinfo
  {volume} {36}},\ \bibinfo {pages} {1113--1116} (\bibinfo {year}
  {1976})}\BibitemShut {NoStop}%
\bibitem [{\citenamefont {Agakishiev}\ \emph {et~al.}(2014)\citenamefont
  {Agakishiev} \emph {et~al.}}]{Agakishiev:2014kdy}%
  \BibitemOpen
  \bibfield  {author} {\bibinfo {author} {\bibfnamefont {G.}~\bibnamefont
  {Agakishiev}} \emph {et~al.} (\bibinfo {collaboration} {HADES}),\ }\bibfield
  {title} {\enquote {\bibinfo {title} {{Lambda hyperon production and
  polarization in collisions of p(3.5 GeV)+Nb}},}\ }\href {\doibase
  10.1140/epja/i2014-14081-2} {\bibfield  {journal} {\bibinfo  {journal} {Eur.
  Phys. J. A}\ }\textbf {\bibinfo {volume} {50}},\ \bibinfo {pages} {81}
  (\bibinfo {year} {2014})},\ \Eprint {http://arxiv.org/abs/1404.3014}
  {arXiv:1404.3014 [nucl-ex]} \BibitemShut {NoStop}%
\bibitem [{\citenamefont {Abt}\ \emph {et~al.}(2006)\citenamefont {Abt} \emph
  {et~al.}}]{Abt:2006da}%
  \BibitemOpen
  \bibfield  {author} {\bibinfo {author} {\bibfnamefont {I.}~\bibnamefont
  {Abt}} \emph {et~al.} (\bibinfo {collaboration} {HERA-B}),\ }\bibfield
  {title} {\enquote {\bibinfo {title} {{Polarization of Lambda and anti-Lambda
  in 920-GeV fixed-target proton-nucleus collisions}},}\ }\href {\doibase
  10.1016/j.physletb.2006.05.040} {\bibfield  {journal} {\bibinfo  {journal}
  {Phys. Lett. B}\ }\textbf {\bibinfo {volume} {638}},\ \bibinfo {pages}
  {415--421} (\bibinfo {year} {2006})},\ \Eprint
  {http://arxiv.org/abs/hep-ex/0603047} {arXiv:hep-ex/0603047} \BibitemShut
  {NoStop}%
\bibitem [{\citenamefont {Becattini}\ \emph {et~al.}(2013)\citenamefont
  {Becattini}, \citenamefont {Chandra}, \citenamefont {Del~Zanna},\ and\
  \citenamefont {Grossi}}]{Becattini:2013fla}%
  \BibitemOpen
  \bibfield  {author} {\bibinfo {author} {\bibfnamefont {F.}~\bibnamefont
  {Becattini}}, \bibinfo {author} {\bibfnamefont {V.}~\bibnamefont {Chandra}},
  \bibinfo {author} {\bibfnamefont {L.}~\bibnamefont {Del~Zanna}}, \ and\
  \bibinfo {author} {\bibfnamefont {E.}~\bibnamefont {Grossi}},\ }\bibfield
  {title} {\enquote {\bibinfo {title} {{Relativistic distribution function for
  particles with spin at local thermodynamical equilibrium}},}\ }\href
  {\doibase 10.1016/j.aop.2013.07.004} {\bibfield  {journal} {\bibinfo
  {journal} {Annals Phys.}\ }\textbf {\bibinfo {volume} {338}},\ \bibinfo
  {pages} {32--49} (\bibinfo {year} {2013})},\ \Eprint
  {http://arxiv.org/abs/1303.3431} {arXiv:1303.3431 [nucl-th]} \BibitemShut
  {NoStop}%
\bibitem [{\citenamefont {Becattini}\ \emph {et~al.}(2017)\citenamefont
  {Becattini}, \citenamefont {Karpenko}, \citenamefont {Lisa}, \citenamefont
  {Upsal},\ and\ \citenamefont {Voloshin}}]{Becattini:2016gvu}%
  \BibitemOpen
  \bibfield  {author} {\bibinfo {author} {\bibfnamefont {F.}~\bibnamefont
  {Becattini}}, \bibinfo {author} {\bibfnamefont {I.}~\bibnamefont {Karpenko}},
  \bibinfo {author} {\bibfnamefont {M.}~\bibnamefont {Lisa}}, \bibinfo {author}
  {\bibfnamefont {I.}~\bibnamefont {Upsal}}, \ and\ \bibinfo {author}
  {\bibfnamefont {S.}~\bibnamefont {Voloshin}},\ }\bibfield  {title} {\enquote
  {\bibinfo {title} {{Global hyperon polarization at local thermodynamic
  equilibrium with vorticity, magnetic field and feed-down}},}\ }\href
  {\doibase 10.1103/PhysRevC.95.054902} {\bibfield  {journal} {\bibinfo
  {journal} {Phys. Rev. C}\ }\textbf {\bibinfo {volume} {95}},\ \bibinfo
  {pages} {054902} (\bibinfo {year} {2017})},\ \Eprint
  {http://arxiv.org/abs/1610.02506} {arXiv:1610.02506 [nucl-th]} \BibitemShut
  {NoStop}%
\bibitem [{\citenamefont {Becattini}\ and\ \citenamefont
  {Lisa}(2020)}]{Becattini:2020ngo}%
  \BibitemOpen
  \bibfield  {author} {\bibinfo {author} {\bibfnamefont {Francesco}\
  \bibnamefont {Becattini}}\ and\ \bibinfo {author} {\bibfnamefont
  {Michael~A.}\ \bibnamefont {Lisa}},\ }\bibfield  {title} {\enquote {\bibinfo
  {title} {{Polarization and Vorticity in the Quark\textendash{}Gluon
  Plasma}},}\ }\href {\doibase 10.1146/annurev-nucl-021920-095245} {\bibfield
  {journal} {\bibinfo  {journal} {Ann. Rev. Nucl. Part. Sci.}\ }\textbf
  {\bibinfo {volume} {70}},\ \bibinfo {pages} {395--423} (\bibinfo {year}
  {2020})},\ \Eprint {http://arxiv.org/abs/2003.03640} {arXiv:2003.03640
  [nucl-ex]} \BibitemShut {NoStop}%
\bibitem [{\citenamefont {Shen}\ and\ \citenamefont
  {Alzhrani}(2020)}]{Shen:2020jwv}%
  \BibitemOpen
  \bibfield  {author} {\bibinfo {author} {\bibfnamefont {Chun}\ \bibnamefont
  {Shen}}\ and\ \bibinfo {author} {\bibfnamefont {Sahr}\ \bibnamefont
  {Alzhrani}},\ }\bibfield  {title} {\enquote {\bibinfo {title}
  {{Collision-geometry-based 3D initial condition for relativistic heavy-ion
  collisions}},}\ }\href {\doibase 10.1103/PhysRevC.102.014909} {\bibfield
  {journal} {\bibinfo  {journal} {Phys. Rev. C}\ }\textbf {\bibinfo {volume}
  {102}},\ \bibinfo {pages} {014909} (\bibinfo {year} {2020})},\ \Eprint
  {http://arxiv.org/abs/2003.05852} {arXiv:2003.05852 [nucl-th]} \BibitemShut
  {NoStop}%
\bibitem [{\citenamefont {Ryu}\ \emph {et~al.}(2021)\citenamefont {Ryu},
  \citenamefont {Jupic},\ and\ \citenamefont {Shen}}]{Ryu:2021lnx}%
  \BibitemOpen
  \bibfield  {author} {\bibinfo {author} {\bibfnamefont {Sangwook}\
  \bibnamefont {Ryu}}, \bibinfo {author} {\bibfnamefont {Vahidin}\ \bibnamefont
  {Jupic}}, \ and\ \bibinfo {author} {\bibfnamefont {Chun}\ \bibnamefont
  {Shen}},\ }\bibfield  {title} {\enquote {\bibinfo {title} {{Probing
  early-time longitudinal dynamics with the \ensuremath{\Lambda} hyperon's spin
  polarization in relativistic heavy-ion collisions}},}\ }\href {\doibase
  10.1103/PhysRevC.104.054908} {\bibfield  {journal} {\bibinfo  {journal}
  {Phys. Rev. C}\ }\textbf {\bibinfo {volume} {104}},\ \bibinfo {pages}
  {054908} (\bibinfo {year} {2021})},\ \Eprint
  {http://arxiv.org/abs/2106.08125} {arXiv:2106.08125 [nucl-th]} \BibitemShut
  {NoStop}%
\bibitem [{\citenamefont {Alzhrani}\ \emph {et~al.}(2022)\citenamefont
  {Alzhrani}, \citenamefont {Ryu},\ and\ \citenamefont
  {Shen}}]{Alzhrani:2022dpi}%
  \BibitemOpen
  \bibfield  {author} {\bibinfo {author} {\bibfnamefont {Sahr}\ \bibnamefont
  {Alzhrani}}, \bibinfo {author} {\bibfnamefont {Sangwook}\ \bibnamefont
  {Ryu}}, \ and\ \bibinfo {author} {\bibfnamefont {Chun}\ \bibnamefont
  {Shen}},\ }\bibfield  {title} {\enquote {\bibinfo {title}
  {{\ensuremath{\Lambda} spin polarization in event-by-event relativistic
  heavy-ion collisions}},}\ }\href {\doibase 10.1103/PhysRevC.106.014905}
  {\bibfield  {journal} {\bibinfo  {journal} {Phys. Rev. C}\ }\textbf {\bibinfo
  {volume} {106}},\ \bibinfo {pages} {014905} (\bibinfo {year} {2022})},\
  \Eprint {http://arxiv.org/abs/2203.15718} {arXiv:2203.15718 [nucl-th]}
  \BibitemShut {NoStop}%
\bibitem [{\citenamefont {Monnai}\ \emph {et~al.}(2019)\citenamefont {Monnai},
  \citenamefont {Schenke},\ and\ \citenamefont {Shen}}]{Monnai:2019hkn}%
  \BibitemOpen
  \bibfield  {author} {\bibinfo {author} {\bibfnamefont {Akihiko}\ \bibnamefont
  {Monnai}}, \bibinfo {author} {\bibfnamefont {Bj\"orn}\ \bibnamefont
  {Schenke}}, \ and\ \bibinfo {author} {\bibfnamefont {Chun}\ \bibnamefont
  {Shen}},\ }\bibfield  {title} {\enquote {\bibinfo {title} {{Equation of state
  at finite densities for QCD matter in nuclear collisions}},}\ }\href
  {\doibase 10.1103/PhysRevC.100.024907} {\bibfield  {journal} {\bibinfo
  {journal} {Phys. Rev. C}\ }\textbf {\bibinfo {volume} {100}},\ \bibinfo
  {pages} {024907} (\bibinfo {year} {2019})},\ \Eprint
  {http://arxiv.org/abs/1902.05095} {arXiv:1902.05095 [nucl-th]} \BibitemShut
  {NoStop}%
\bibitem [{\citenamefont {Schenke}\ \emph {et~al.}(2010)\citenamefont
  {Schenke}, \citenamefont {Jeon},\ and\ \citenamefont
  {Gale}}]{Schenke:2010nt}%
  \BibitemOpen
  \bibfield  {author} {\bibinfo {author} {\bibfnamefont {Bjoern}\ \bibnamefont
  {Schenke}}, \bibinfo {author} {\bibfnamefont {Sangyong}\ \bibnamefont
  {Jeon}}, \ and\ \bibinfo {author} {\bibfnamefont {Charles}\ \bibnamefont
  {Gale}},\ }\bibfield  {title} {\enquote {\bibinfo {title} {{(3+1)D
  hydrodynamic simulation of relativistic heavy-ion collisions}},}\ }\href
  {\doibase 10.1103/PhysRevC.82.014903} {\bibfield  {journal} {\bibinfo
  {journal} {Phys. Rev. C}\ }\textbf {\bibinfo {volume} {82}},\ \bibinfo
  {pages} {014903} (\bibinfo {year} {2010})},\ \Eprint
  {http://arxiv.org/abs/1004.1408} {arXiv:1004.1408 [hep-ph]} \BibitemShut
  {NoStop}%
\bibitem [{\citenamefont {Schenke}\ \emph {et~al.}(2011)\citenamefont
  {Schenke}, \citenamefont {Jeon},\ and\ \citenamefont
  {Gale}}]{Schenke:2010rr}%
  \BibitemOpen
  \bibfield  {author} {\bibinfo {author} {\bibfnamefont {Bjorn}\ \bibnamefont
  {Schenke}}, \bibinfo {author} {\bibfnamefont {Sangyong}\ \bibnamefont
  {Jeon}}, \ and\ \bibinfo {author} {\bibfnamefont {Charles}\ \bibnamefont
  {Gale}},\ }\bibfield  {title} {\enquote {\bibinfo {title} {{Elliptic and
  triangular flow in event-by-event (3+1)D viscous hydrodynamics}},}\ }\href
  {\doibase 10.1103/PhysRevLett.106.042301} {\bibfield  {journal} {\bibinfo
  {journal} {Phys. Rev. Lett.}\ }\textbf {\bibinfo {volume} {106}},\ \bibinfo
  {pages} {042301} (\bibinfo {year} {2011})},\ \Eprint
  {http://arxiv.org/abs/1009.3244} {arXiv:1009.3244 [hep-ph]} \BibitemShut
  {NoStop}%
\bibitem [{\citenamefont {Paquet}\ \emph {et~al.}(2016)\citenamefont {Paquet},
  \citenamefont {Shen}, \citenamefont {Denicol}, \citenamefont {Luzum},
  \citenamefont {Schenke}, \citenamefont {Jeon},\ and\ \citenamefont
  {Gale}}]{Paquet:2015lta}%
  \BibitemOpen
  \bibfield  {author} {\bibinfo {author} {\bibfnamefont {Jean-Fran\c{c}ois}\
  \bibnamefont {Paquet}}, \bibinfo {author} {\bibfnamefont {Chun}\ \bibnamefont
  {Shen}}, \bibinfo {author} {\bibfnamefont {Gabriel~S.}\ \bibnamefont
  {Denicol}}, \bibinfo {author} {\bibfnamefont {Matthew}\ \bibnamefont
  {Luzum}}, \bibinfo {author} {\bibfnamefont {Bj\"orn}\ \bibnamefont
  {Schenke}}, \bibinfo {author} {\bibfnamefont {Sangyong}\ \bibnamefont
  {Jeon}}, \ and\ \bibinfo {author} {\bibfnamefont {Charles}\ \bibnamefont
  {Gale}},\ }\bibfield  {title} {\enquote {\bibinfo {title} {{Production of
  photons in relativistic heavy-ion collisions}},}\ }\href {\doibase
  10.1103/PhysRevC.93.044906} {\bibfield  {journal} {\bibinfo  {journal} {Phys.
  Rev. C}\ }\textbf {\bibinfo {volume} {93}},\ \bibinfo {pages} {044906}
  (\bibinfo {year} {2016})},\ \Eprint {http://arxiv.org/abs/1509.06738}
  {arXiv:1509.06738 [hep-ph]} \BibitemShut {NoStop}%
\bibitem [{\citenamefont {Denicol}\ \emph {et~al.}(2018)\citenamefont
  {Denicol}, \citenamefont {Gale}, \citenamefont {Jeon}, \citenamefont
  {Monnai}, \citenamefont {Schenke},\ and\ \citenamefont
  {Shen}}]{Denicol:2018wdp}%
  \BibitemOpen
  \bibfield  {author} {\bibinfo {author} {\bibfnamefont {Gabriel~S.}\
  \bibnamefont {Denicol}}, \bibinfo {author} {\bibfnamefont {Charles}\
  \bibnamefont {Gale}}, \bibinfo {author} {\bibfnamefont {Sangyong}\
  \bibnamefont {Jeon}}, \bibinfo {author} {\bibfnamefont {Akihiko}\
  \bibnamefont {Monnai}}, \bibinfo {author} {\bibfnamefont {Bj\"orn}\
  \bibnamefont {Schenke}}, \ and\ \bibinfo {author} {\bibfnamefont {Chun}\
  \bibnamefont {Shen}},\ }\bibfield  {title} {\enquote {\bibinfo {title} {{Net
  baryon diffusion in fluid dynamic simulations of relativistic heavy-ion
  collisions}},}\ }\href {\doibase 10.1103/PhysRevC.98.034916} {\bibfield
  {journal} {\bibinfo  {journal} {Phys. Rev. C}\ }\textbf {\bibinfo {volume}
  {98}},\ \bibinfo {pages} {034916} (\bibinfo {year} {2018})},\ \Eprint
  {http://arxiv.org/abs/1804.10557} {arXiv:1804.10557 [nucl-th]} \BibitemShut
  {NoStop}%
\bibitem [{\citenamefont {Denicol}\ \emph {et~al.}(2012)\citenamefont
  {Denicol}, \citenamefont {Niemi}, \citenamefont {Molnar},\ and\ \citenamefont
  {Rischke}}]{Denicol:2012cn}%
  \BibitemOpen
  \bibfield  {author} {\bibinfo {author} {\bibfnamefont {G.~S.}\ \bibnamefont
  {Denicol}}, \bibinfo {author} {\bibfnamefont {H.}~\bibnamefont {Niemi}},
  \bibinfo {author} {\bibfnamefont {E.}~\bibnamefont {Molnar}}, \ and\ \bibinfo
  {author} {\bibfnamefont {D.~H.}\ \bibnamefont {Rischke}},\ }\bibfield
  {title} {\enquote {\bibinfo {title} {{Derivation of transient relativistic
  fluid dynamics from the Boltzmann equation}},}\ }\href {\doibase
  10.1103/PhysRevD.85.114047} {\bibfield  {journal} {\bibinfo  {journal} {Phys.
  Rev. D}\ }\textbf {\bibinfo {volume} {85}},\ \bibinfo {pages} {114047}
  (\bibinfo {year} {2012})},\ \bibinfo {note} {[Erratum: Phys.Rev.D 91, 039902
  (2015)]},\ \Eprint {http://arxiv.org/abs/1202.4551} {arXiv:1202.4551
  [nucl-th]} \BibitemShut {NoStop}%
\bibitem [{\citenamefont {Huovinen}\ and\ \citenamefont
  {Petersen}(2012)}]{Huovinen:2012is}%
  \BibitemOpen
  \bibfield  {author} {\bibinfo {author} {\bibfnamefont {Pasi}\ \bibnamefont
  {Huovinen}}\ and\ \bibinfo {author} {\bibfnamefont {Hannah}\ \bibnamefont
  {Petersen}},\ }\bibfield  {title} {\enquote {\bibinfo {title}
  {{Particlization in hybrid models}},}\ }\href {\doibase
  10.1140/epja/i2012-12171-9} {\bibfield  {journal} {\bibinfo  {journal} {Eur.
  Phys. J. A}\ }\textbf {\bibinfo {volume} {48}},\ \bibinfo {pages} {171}
  (\bibinfo {year} {2012})},\ \Eprint {http://arxiv.org/abs/1206.3371}
  {arXiv:1206.3371 [nucl-th]} \BibitemShut {NoStop}%
\bibitem [{\citenamefont {Adare}\ \emph {et~al.}(2018)\citenamefont {Adare}
  \emph {et~al.}}]{PHENIX:2018hho}%
  \BibitemOpen
  \bibfield  {author} {\bibinfo {author} {\bibfnamefont {A.}~\bibnamefont
  {Adare}} \emph {et~al.} (\bibinfo {collaboration} {PHENIX}),\ }\bibfield
  {title} {\enquote {\bibinfo {title} {{Pseudorapidity Dependence of Particle
  Production and Elliptic Flow in Asymmetric Nuclear Collisions of $p+$Al,
  $p+$Au, $d+$Au, and $^{3}$He$+$Au at $\sqrt{s_{_{NN}}}=200$ GeV}},}\ }\href
  {\doibase 10.1103/PhysRevLett.121.222301} {\bibfield  {journal} {\bibinfo
  {journal} {Phys. Rev. Lett.}\ }\textbf {\bibinfo {volume} {121}},\ \bibinfo
  {pages} {222301} (\bibinfo {year} {2018})},\ \Eprint
  {http://arxiv.org/abs/1807.11928} {arXiv:1807.11928 [nucl-ex]} \BibitemShut
  {NoStop}%
\bibitem [{\citenamefont {Back}\ \emph {et~al.}(2006)\citenamefont {Back} \emph
  {et~al.}}]{PHOBOS:2005zhy}%
  \BibitemOpen
  \bibfield  {author} {\bibinfo {author} {\bibfnamefont {B.~B.}\ \bibnamefont
  {Back}} \emph {et~al.} (\bibinfo {collaboration} {PHOBOS}),\ }\bibfield
  {title} {\enquote {\bibinfo {title} {{Charged-particle pseudorapidity
  distributions in Au+Au collisions at $s(NN) ^{1/2}$ = 62.4-GeV}},}\ }\href
  {\doibase 10.1103/PhysRevC.74.021901} {\bibfield  {journal} {\bibinfo
  {journal} {Phys. Rev. C}\ }\textbf {\bibinfo {volume} {74}},\ \bibinfo
  {pages} {021901} (\bibinfo {year} {2006})},\ \Eprint
  {http://arxiv.org/abs/nucl-ex/0509034} {arXiv:nucl-ex/0509034} \BibitemShut
  {NoStop}%
\bibitem [{\citenamefont {Hidaka}\ \emph {et~al.}(2018)\citenamefont {Hidaka},
  \citenamefont {Pu},\ and\ \citenamefont {Yang}}]{Hidaka:2017auj}%
  \BibitemOpen
  \bibfield  {author} {\bibinfo {author} {\bibfnamefont {Yoshimasa}\
  \bibnamefont {Hidaka}}, \bibinfo {author} {\bibfnamefont {Shi}\ \bibnamefont
  {Pu}}, \ and\ \bibinfo {author} {\bibfnamefont {Di-Lun}\ \bibnamefont
  {Yang}},\ }\bibfield  {title} {\enquote {\bibinfo {title} {{Nonlinear
  Responses of Chiral Fluids from Kinetic Theory}},}\ }\href {\doibase
  10.1103/PhysRevD.97.016004} {\bibfield  {journal} {\bibinfo  {journal} {Phys.
  Rev. D}\ }\textbf {\bibinfo {volume} {97}},\ \bibinfo {pages} {016004}
  (\bibinfo {year} {2018})},\ \Eprint {http://arxiv.org/abs/1710.00278}
  {arXiv:1710.00278 [hep-th]} \BibitemShut {NoStop}%
\bibitem [{\citenamefont {Liu}\ and\ \citenamefont
  {Yin}(2021{\natexlab{a}})}]{Liu:2020dxg}%
  \BibitemOpen
  \bibfield  {author} {\bibinfo {author} {\bibfnamefont {Shuai Y.~F.}\
  \bibnamefont {Liu}}\ and\ \bibinfo {author} {\bibfnamefont {Yi}~\bibnamefont
  {Yin}},\ }\bibfield  {title} {\enquote {\bibinfo {title} {{Spin Hall effect
  in heavy-ion collisions}},}\ }\href {\doibase 10.1103/PhysRevD.104.054043}
  {\bibfield  {journal} {\bibinfo  {journal} {Phys. Rev. D}\ }\textbf {\bibinfo
  {volume} {104}},\ \bibinfo {pages} {054043} (\bibinfo {year}
  {2021}{\natexlab{a}})},\ \Eprint {http://arxiv.org/abs/2006.12421}
  {arXiv:2006.12421 [nucl-th]} \BibitemShut {NoStop}%
\bibitem [{\citenamefont {Liu}\ and\ \citenamefont
  {Yin}(2021{\natexlab{b}})}]{Liu:2021uhn}%
  \BibitemOpen
  \bibfield  {author} {\bibinfo {author} {\bibfnamefont {Shuai Y.~F.}\
  \bibnamefont {Liu}}\ and\ \bibinfo {author} {\bibfnamefont {Yi}~\bibnamefont
  {Yin}},\ }\bibfield  {title} {\enquote {\bibinfo {title} {{Spin polarization
  induced by the hydrodynamic gradients}},}\ }\href {\doibase
  10.1007/JHEP07(2021)188} {\bibfield  {journal} {\bibinfo  {journal} {JHEP}\
  }\textbf {\bibinfo {volume} {07}},\ \bibinfo {pages} {188} (\bibinfo {year}
  {2021}{\natexlab{b}})},\ \Eprint {http://arxiv.org/abs/2103.09200}
  {arXiv:2103.09200 [hep-ph]} \BibitemShut {NoStop}%
\bibitem [{\citenamefont {Becattini}\ \emph {et~al.}(2021)\citenamefont
  {Becattini}, \citenamefont {Buzzegoli},\ and\ \citenamefont
  {Palermo}}]{Becattini:2021suc}%
  \BibitemOpen
  \bibfield  {author} {\bibinfo {author} {\bibfnamefont {F.}~\bibnamefont
  {Becattini}}, \bibinfo {author} {\bibfnamefont {M.}~\bibnamefont
  {Buzzegoli}}, \ and\ \bibinfo {author} {\bibfnamefont {A.}~\bibnamefont
  {Palermo}},\ }\bibfield  {title} {\enquote {\bibinfo {title} {{Spin-thermal
  shear coupling in a relativistic fluid}},}\ }\href {\doibase
  10.1016/j.physletb.2021.136519} {\bibfield  {journal} {\bibinfo  {journal}
  {Phys. Lett. B}\ }\textbf {\bibinfo {volume} {820}},\ \bibinfo {pages}
  {136519} (\bibinfo {year} {2021})},\ \Eprint
  {http://arxiv.org/abs/2103.10917} {arXiv:2103.10917 [nucl-th]} \BibitemShut
  {NoStop}%
\bibitem [{\citenamefont {Buzzegoli}(2022)}]{Buzzegoli:2021wlg}%
  \BibitemOpen
  \bibfield  {author} {\bibinfo {author} {\bibfnamefont {M.}~\bibnamefont
  {Buzzegoli}},\ }\bibfield  {title} {\enquote {\bibinfo {title} {{Pseudogauge
  dependence of the spin polarization and of the axial vortical effect}},}\
  }\href {\doibase 10.1103/PhysRevC.105.044907} {\bibfield  {journal} {\bibinfo
   {journal} {Phys. Rev. C}\ }\textbf {\bibinfo {volume} {105}},\ \bibinfo
  {pages} {044907} (\bibinfo {year} {2022})},\ \Eprint
  {http://arxiv.org/abs/2109.12084} {arXiv:2109.12084 [nucl-th]} \BibitemShut
  {NoStop}%
\bibitem [{\citenamefont {Liu}\ and\ \citenamefont
  {Huang}(2022)}]{Liu:2021nyg}%
  \BibitemOpen
  \bibfield  {author} {\bibinfo {author} {\bibfnamefont {Yu-Chen}\ \bibnamefont
  {Liu}}\ and\ \bibinfo {author} {\bibfnamefont {Xu-Guang}\ \bibnamefont
  {Huang}},\ }\bibfield  {title} {\enquote {\bibinfo {title} {{Spin
  polarization formula for Dirac fermions at local equilibrium}},}\ }\href
  {\doibase 10.1007/s11433-022-1903-8} {\bibfield  {journal} {\bibinfo
  {journal} {Sci. China Phys. Mech. Astron.}\ }\textbf {\bibinfo {volume}
  {65}},\ \bibinfo {pages} {272011} (\bibinfo {year} {2022})},\ \Eprint
  {http://arxiv.org/abs/2109.15301} {arXiv:2109.15301 [nucl-th]} \BibitemShut
  {NoStop}%
\bibitem [{\citenamefont {Pordes}\ \emph {et~al.}(2007)\citenamefont {Pordes}
  \emph {et~al.}}]{Pordes:2007zzb}%
  \BibitemOpen
  \bibfield  {author} {\bibinfo {author} {\bibfnamefont {Ruth}\ \bibnamefont
  {Pordes}} \emph {et~al.},\ }\bibfield  {title} {\enquote {\bibinfo {title}
  {{The Open Science Grid}},}\ }\href {\doibase 10.1088/1742-6596/78/1/012057}
  {\bibfield  {journal} {\bibinfo  {journal} {J. Phys. Conf. Ser.}\ }\textbf
  {\bibinfo {volume} {78}},\ \bibinfo {pages} {012057} (\bibinfo {year}
  {2007})}\BibitemShut {NoStop}%
\bibitem [{\citenamefont {Sfiligoi}\ \emph {et~al.}(2009)\citenamefont
  {Sfiligoi}, \citenamefont {Bradley}, \citenamefont {Holzman}, \citenamefont
  {Mhashilkar}, \citenamefont {Padhi},\ and\ \citenamefont
  {Wurthwrin}}]{Sfiligoi:2009cct}%
  \BibitemOpen
  \bibfield  {author} {\bibinfo {author} {\bibfnamefont {Igor}\ \bibnamefont
  {Sfiligoi}}, \bibinfo {author} {\bibfnamefont {Daniel~C.}\ \bibnamefont
  {Bradley}}, \bibinfo {author} {\bibfnamefont {Burt}\ \bibnamefont {Holzman}},
  \bibinfo {author} {\bibfnamefont {Parag}\ \bibnamefont {Mhashilkar}},
  \bibinfo {author} {\bibfnamefont {Sanjay}\ \bibnamefont {Padhi}}, \ and\
  \bibinfo {author} {\bibfnamefont {Frank}\ \bibnamefont {Wurthwrin}},\
  }\bibfield  {title} {\enquote {\bibinfo {title} {{The pilot way to Grid
  resources using glideinWMS}},}\ }\href {\doibase 10.1109/CSIE.2009.950}
  {\bibfield  {journal} {\bibinfo  {journal} {WRI World Congress}\ }\textbf
  {\bibinfo {volume} {2}},\ \bibinfo {pages} {428--432} (\bibinfo {year}
  {2009})}\BibitemShut {NoStop}%
\bibitem [{\citenamefont {{OSG}}(2006)}]{https://doi.org/10.21231/906p-4d78}%
  \BibitemOpen
  \bibfield  {author} {\bibinfo {author} {\bibnamefont {{OSG}}},\ }\href
  {\doibase 10.21231/906P-4D78} {\enquote {\bibinfo {title} {Ospool},}\ }
  (\bibinfo {year} {2006})\BibitemShut {NoStop}%
\bibitem [{\citenamefont {{OSG}}(2015)}]{https://doi.org/10.21231/0kvz-ve57}%
  \BibitemOpen
  \bibfield  {author} {\bibinfo {author} {\bibnamefont {{OSG}}},\ }\href
  {\doibase 10.21231/0KVZ-VE57} {\enquote {\bibinfo {title} {Open science data
  federation},}\ } (\bibinfo {year} {2015})\BibitemShut {NoStop}%
\end{thebibliography}%

\end{document}